\documentclass[twocolumn]{aastex7_cranmer2026}
\usepackage{hyperref}
\usepackage{times}

\shorttitle{Testing Theories of Coronal Heating}
\shortauthors{S.\  R.\  Cranmer and C.\  R.\  Gilly}

\begin{document}

\title{Testing Theories of Solar Coronal Heating with
Three-Dimensional Forward Modeling}

\correspondingauthor{Steven R. Cranmer}
\author[orcid=0000-0002-3699-3134]{Steven R. Cranmer}
\affiliation{Department of Astrophysical and Planetary Sciences,
Laboratory for Atmospheric and Space Physics,
University of Colorado, Boulder, CO 80309, USA}
\email[show]{steven.cranmer@colorado.edu}

\author[orcid=0000-0003-0021-9056]{Chris R. Gilly}
\affiliation{NorthWest Research Associates, Boulder, CO 80301, USA}
\email[show]{gilly@nwra.com}

\begin{abstract}
After almost a century of speculation, the physical processes
responsible for heating the Sun's corona remain uncertain.
Observations of coronal loops at extreme ultraviolet and X-ray wavelengths
have provided substantial insights, such as the constraint that
heating rates increase monotonically with local magnetic field strength.
However, there remain dozens of competing suggestions for detailed
theoretical explanations of the heating.
There remains a need to compare the predictions of these theories with
one another, and with real data, so the most likely mechanisms
can be determined.
In this paper, we build three-dimensional simulations of the corona
by starting with potential-field extrapolations for the magnetic field,
filling in plasma densities and temperatures along individual loops
using one-dimensional models, then computing optically thin intensities
for comparison with data from the Solar Dynamics Observatory and Hinode.
One new aspect of this forward modeling is the use of parameterizations
for heating that include a wide variety of processes such as
wave dissipation, reconnection, nanoflares, and turbulent cascade.
We also include the effects of intermittent heating and cooling
along unresolved strands in the form of time-averaged multithermal
broadening of the local differential emission measure.
After allowing the coronal heating prescription to vary freely, we found
a narrow range of parameters that produce optimal agreement with
a set of more than 500 independently measured intensities at solar
minimum and maximum.
The values of these parameters are consistent with those predicted
by models of imbalanced magnetohydrodynamic turbulence, but this does not
necessarily exclude contributions from other heating processes.
\end{abstract}


\section{Introduction}
\label{sec:intro}

Many different processes have been suggested to explain the
origin of the corona and solar wind, but we still have not
conclusively identified the detailed pathways of energy transfer
from convection and magnetic fields to plasma heating.
Understanding and characterizing these processes is important not only
for improving forecasts of potentially dangerous space weather events
\citep{Ks17,Tm21},
but also for establishing a baseline of knowledge about a well-resolved
system that is directly relevant to other astrophysical environments.
There has been substantial progress in producing three-dimensional
simulations that reproduce many observed properties of the corona, but
these tools are often limited by coarse spatial and time
resolution that does not resolve granulation-scale dynamics
\citep[e.g.,][]{Pe15} or by including only a single proposed
heating mechanism, such as magnetohydrodynamic (MHD) turbulence
\citep[e.g.,][]{Go18}.

In general, there are two families of methods for deciding whether a
given theoretical prediction agrees with a set of observational data:
(1) inverse modeling, in which the data are used to infer the most likely
physical parameters (i.e., densities, temperatures, velocities,
and elemental abundances); and
(2) forward modeling, in which theory is used to simulate the
three-dimensional plasma state, then that is used to synthesize
what an observer would see.
One-to-one comparisons are performed at either the level of the plasma
parameters (for inverse models) or the observables (for forward models).
A broad range of inverse methods are used in solar physics
\citep[e.g.,][]{KD98,HK12,Ln12,DZ13,Nu13,As15,De22,Mc22,Pj23}.
However, because the corona is optically thin and highly variable in space
and time, some attempts at data inversion are nonunique and susceptible
to significant uncertainties (see many of the papers listed above;
additionally, \citeauthor{CB76} \citeyear{CB76};
\citeauthor{Ju10} \citeyear{Ju10}).

For many applications, the forward-modeling approach is understood as a
safer route, and a large number of tools have been developed to allow for
synthetic observables to be constructed
\citep{Mok05,Mok16,Gi16,VD16,Mik18,Nita18,Sz19,Fy21,Pt22,Nita23,Breu25,Ri25}.
This paper describes new results from a type of forward modeling that
has been used in earlier studies of both the Sun and other stars.
In these studies
\citep[e.g.,][]{KP95,Mn00,Sj04,WW06,Lu08,Wi08,DD11,Fd17,Ug19,Fs21,Is24},
the coronal magnetic field is reconstructed using
extrapolation from measurements of the photospheric field.
The coronal heating rate is then parameterized in terms of analytic
dependences on local quantities like the magnetic field $B$ and length $L$
of a given coronal loop.
Thus, the magnetic field is treated as a known ``skeleton'' and
one-dimensional coronal loop models are used to apply the ``musculature''
(i.e., plasma densities and temperatures).
Because there is no discrete numerical grid in the corona, the spatial
resolution for loops and other field-aligned strands is not fixed in
advance and can be made as fine as desired.

In this paper, we choose to parameterize the coronal heating rate using
expressions that encompass a wide variety of proposed heating processes
such as wave dissipation, reconnection, nanoflares, and turbulent cascade.
We also include the effects of ambipolar diffusion on electron
heat conduction in the transition region, and we account for
multithermal broadening of the differential emission measure
distribution due to intermittent nanoflare-like energy release events.
To find the model parameters that optimally match observations, we
simulate dozens of optically thin lines of sight through the corona
during solar minimum and maximum epochs, then compare with
extreme ultraviolet (EUV) and X-ray data.

Section~\ref{sec:heating} describes our approach to specifying
the coronal heating rate.
In Section~\ref{sec:plasma}, we present the other ingredients of
the forward model, such as the extrapolated magnetic field, thermal
conduction, radiative losses, time-variability effects, and our solution
method for the time-steady plasma properties.
Section~\ref{sec:emission} then discusses the data---EUV images from the
Solar Dynamics Observatory (SDO) and X-ray images from the Hinode
spacecraft---which we use to compare with the synthesized observables.
Some initial and idealized results of the modeling, which help us
understand how various parameters affect the simulated
properties of the corona, are described in Section~\ref{sec:params}.
Thereafter, a more detailed comparison between the model predictions and
the observations is provided in Section~\ref{sec:results}.
Lastly, Section~\ref{sec:conc} concludes with a brief summary, a
discussion of some of the wider implications of these results, and
suggestions for future improvements.

\section{Coronal Heating Models}
\label{sec:heating}

For magnetic field lines with no large-scale plasma flows,
we treat the time-steady energization of the chromosphere and corona
as a balance between three spatially dependent rates of heating or
cooling, with
\begin{equation}
  Q_{\rm heat} + Q_{\rm cond} + Q_{\rm rad} \, = \, 0 \,\, .
  \label{eq:Qbal}
\end{equation}
and the three terms on the left-hand side are associated with
direct heating, thermal conduction, and coupling to the radiation field,
respectively.
These ``volumetric'' $Q$ rates have units of power per unit volume,
so they can be considered either as an energy density gained or lost
per unit time, or as an energy flux distributed over a finite length.
The conduction and radiation terms are discussed in more detail in
Section~\ref{sec:plasma}.

\subsection{Scaling with Poynting Flux}
\label{sec:heating:mine}

In this paper, we choose to work with an effectively time-steady
heating rate $Q_{\rm heat}$.
For a given coronal loop, this can be interpreted as statistical mean
value that allows for the presence of phenomena such as waves,
intermittent turbulence, or bursty nanoflares.
However, there is still substantial uncertainty about whether an accurate
characterization of coronal heating must involve the sum over many
sporadic and time-variable structures, or whether a mean heating rate
can adequately describe the physics
\citep[e.g.,][]{P88,VK11,VK16,DMB15,Js19,Ju23}.

Many proposed models of coronal heating have rates that depend on
four primary parameters, two that describe the photospheric base
and two that describe the coronal field line in question:
\begin{enumerate}
\item
The transverse velocity amplitude $v_{\perp}$, which is usually assumed
to be a root-mean-squared (rms) quantity, characterizes the amplitude of
horizontal ``footpoint'' motions due to convective granulation.
Typical photospheric values from both observations (e.g., correlation
tracking of granular motions) and simulations tend to be
of order 0.5 to 1 km~s$^{-1}$ \citep{NS88,We15}, with larger values
up to 5 km~s$^{-1}$ occasionally seen at the smallest scales
\citep[e.g.,][]{Bg98}.
We generally assume a fixed fiducial value of 1 km~s$^{-1}$, but we
must also take account of the fact that convective motions are
suppressed in the strongest-field regions exhibiting sunspot or
pore-like dimmings.
Thus, we follow \citet{Sj04} by multiplying the fiducial value of
$v_{\perp}$ by a quenching factor $\exp[-(B_{\odot}/B_{\rm spot})^2]$,
where $B_{\odot}$ is the magnetic field at the coronal base and
$B_{\rm spot} = 500$~G \citep[see also][]{Tw17,Ap25}.
\item
The transverse correlation length $\lambda_{\perp}$ describes the
most likely photospheric spatial scale corresponding to the
$v_{\perp}$ motions.
Typical values chosen for this quantity often span a range from
the size of granules (i.e., 1~Mm; \citeauthor{Ab13} \citeyear{Ab13})
to the size of chromospheric network patches (5--10 Mm) and
supergranular cells (20--30 Mm; \citeauthor{Dm02} \citeyear{Dm02}).
However, high-resolution observations have begun to show that
much of the magnetic field in the chromosphere and low corona is
connected to small tube-like patches of kiloGauss-strength fields
in narrow intergranular lanes \citep[see, e.g.,][]{CvB05}.
Thus, for this paper we set $\lambda_{\perp}$ at a value of 0.120~Mm,
which is of the same order of magnitude as both (a) typical horizontal
scales of intergranular bright points, and (b) the vertical scale
height of the photosphere.
\item
For closed magnetic-field regions, the coronal loop length $L$
helps constrain the timescale over which waves and other
fluctuations propagate from one footpoint to the other.
Specifically, we define $L$ as {\em half} of the total
footpoint-to-footpoint length; i.e., for a symmetric loop it is
the field-aligned distance from one footpoint to its apex.
We describe how open field lines (connected to the solar wind)
are specified below.
Values of $L$ for specific regions of the corona are found
numerically by tracing through three-dimensional extrapolations
of the magnetic field, which is one of the most computationally
intensive parts of this modeling approach.
\item
In the corona, a key characteristic velocity is the local
Alfv\'{e}n speed $V_{\rm A} = B/\sqrt{4\pi\rho}$,
where $B$ is the magnetic field strength at a particular
location and $\rho$ is the mass density there.
The magnetic field strength is obtained from the same extrapolation
used to determine $L$, and the dependence on $\rho$ is
discussed in more detail below.
\end{enumerate}
One intuitive way to parameterize the heating rate is to assume
that it scales with the vertical component of the MHD Poynting flux
associated with horizontal motions at the photosphere \citep{We15},
then to assume this energy flux is distributed over a coronal loop
of length $L$.
This gives a volumetric heating rate of
\begin{equation}
  Q_{\rm heat} \, = \, {\cal E} \,
  \frac{\rho v_{\perp} V_{\rm A}^2}{L}  \,\, ,
  \label{eq:poynt}
\end{equation}
where ${\cal E}$ is a dimensionless efficiency factor.
This expression assumes the Poynting flux is dominated by the
horizontal jostling of magnetic field lines and that its other
components (due to, e.g., flux emergence through the photosphere)
are less important; see, however, \citet{CI14} and \citet{Kn18}.

\citet{CW19} summarized how a large number of proposed coronal heating
theories can be encapsulated by using different power-law dependences
for the efficiency factor:
\begin{equation}
  {\cal E} \, = \, {\cal E}_0
  \left( \frac{\tau_{\rm A}}{\tau_{\rm ph}} \right)^m
  \left( \frac{\lambda_{\perp}}{L} \right)^n
  \label{eq:efficiency}
\end{equation}
where $\tau_{\rm A} = L / V_{\rm A}$ is a representative
Alfv\'{e}n-wave travel time through the loop and
$\tau_{\rm ph} = \lambda_{\perp} / v_{\perp}$ is a representative
photospheric driving timescale.
Traditionally, scenarios with $\tau_{\rm A} \ll \tau_{\rm ph}$
are compared by analogy with direct-current (DC) electrical systems,
and those with $\tau_{\rm A} \gg \tau_{\rm ph}$
are compared with alternating-current (AC) systems.

\begin{table*}
\caption{Scaling Exponents from Proposed Coronal Heating
Theories and Simulations
\label{table01}}
\hspace*{0.35in}
\begin{tabular}{lccl}
\hline
\hline
Physical Description & $m$ & $n$ & Example References \\
\hline
Current-layer random walk & 0 & 1 &
  \citet{P72} \\
DC braided turbulence & 0 $\rightarrow$ 0.5 & 1 &
  \citet{Rp08} \\
Current-layer shearing & 0 $\rightarrow$ 1 & 0 $\rightarrow$ 1 &
  \citet{GN96} \\
2D boundary-driven cascade & 0.333 & 0.667 &
  \citet{HP92} \\
Line-tied reduced MHD cascade & 0.5 & 1 &
  \citet{DG99} \\
Alfv\'{e}n wave Poynting flux & 1 & 1 &
  \citet{A47} \\
Braided discontinuities & 1 & 2 &
  \citet{P83} \\
Kolmogorov turbulence & 2 & 1 & 
  \citet{dKH38} \\
Hybrid triple-correlation cascade & 2 $\rightarrow$ 3 & 1 &
  \citet{ZM90} \\
Iroshnikov-Kraichnan turbulence & 3 & 2 & 
  \citet{Chae02,Bd06} \\
Switch-on MHD shock train & 4 & 3 &
  \citet{Ho85} \\
\hline
Reduced MHD turbulence & 0.65 $\rightarrow$ 1.45 & 0.57 $\rightarrow$ 1.37 &
  \citet{vB11} \\
3D braided simulation & 0.25 $\rightarrow$ 0.78 & 0.90 $\rightarrow$ 1.43 &
  \citet{Bou16} \\
Reflection-driven turbulence & 0.12 $\rightarrow$ 0.76 & --0.42 $\rightarrow$ 1.18 &
  \citet{Do16} \\
\hline
\end{tabular}
\end{table*}

Table~\ref{table01} provides a selection of predictions
for the $m$ and $n$ exponents from some analytic theories and
computer simulations.
The associated references to the literature are not meant to be
comprehensive; see \citet{CW19} for additional citations for many
of these mechanisms.
For example, the scaling associated with the current-layer random walk
model has been reinvented independently in different contexts
\citep{SU81,vB86,Ng08}.
The entry associated with \citet{A47} refers to a simple model in which
a constant fraction of the Poynting flux associated with linear
Alfv\'{e}n waves is converted into heat.
The entry associated with Kolmogorov turbulence could also apply
to the anisotropic turbulence model of \citet{GS95} or to models of
reflection-driven turbulence \citep[e.g.,][]{Hs95,Mt99} with a
constant imbalance fraction.

In this paper, we generally assume that $v_{\perp}$ and $\lambda_{\perp}$
refer to photospheric driving values.
However, in some coronal heating models these quantities refer to
spatially-dependent fluctuations that vary from the photosphere to
the corona.
Observationally, coronal transverse velocities tend to be about a
factor of 10 to 20 larger than the photospheric values,
and the correlation lengths are often assumed to expand with the
transverse sizes of magnetic flux tubes (i.e.,
$\lambda_{\perp} \propto B^{-1/2}$).
\citet{Rp08} discussed various ways of converting energy fluxes that
originally were expressed in terms of local values into expressions
that depend on the photospheric boundary conditions.

In order to more clearly specify how the heating rate depends on its
parameters, and on spatial position, we rewrite 
Equations~(\ref{eq:poynt})--(\ref{eq:efficiency}) by separating out
the dependences more explicitly as
\begin{equation}
  Q_{\rm heat} \, = \, {\cal E}_0 \,
  v_{\perp}^{m+1} \, \lambda_{\perp}^{n-m} \, L^{m-n-1}
  \left( \frac{B}{\sqrt{4\pi}} \right)^{2-m}
  \rho^{m/2} \,\,\, .
  \label{eq:Qheatvsmn}
\end{equation}
For a given field line, ${\cal E}_0$, $v_{\perp}$, $\lambda_{\perp}$,
and $L$ are essentially constants, but both
$B$ and $\rho$ vary as a function of position.
Let us use $s$ to specify the length coordinate mapped along a given
coronal loop.
Anticipating the use of the solution method of \citet{Ma10}, we also
parameterize the decrease in $B$ as an inverse proportionality with
the rise in temperature,
\begin{equation}
  \frac{B(s)}{B_{\odot}} \, = \,
  \left[ \frac{T_{\rm base}}{T(s)} \right]^{\delta}
  \label{eq:deltadef}
\end{equation}
where $T_{\rm base}$ is fixed at a typical chromospheric value of
$10^4$~K and $B_{\odot}$ is the basal field strength in the photosphere.
Because we limit ourselves to loops with $B(s) \leq B_{\odot}$ and
$T(s) \geq T_{\rm base}$, the exponent $\delta$ is constrained to
be positive.
Using the above parameterizations, the heating rate can be written as
\begin{displaymath}
  Q_{\rm heat} \, = \, \Bigg[ {\cal E}_0 \, v_{\perp}^{m+1} \,
  \lambda_{\perp}^{n-m} \, L^{m-n-1} \,\,\, \times
\end{displaymath}
\begin{equation}
  \times \,\,\, T_{\rm base}^{\delta (2-m)}
  \left( \frac{B_{\odot}}{\sqrt{4\pi}} \right)^{2-m}
  \Bigg] \rho^{m/2} \, T^{\delta (m-2)}
  \label{eq:Qpractical}
\end{equation}
where the quantities inside the square brackets do not depend on $s$,
but the ones outside the brackets do.

Open field lines, with only one footpoint on the Sun's surface,
ought to be handled with a combined model of coronal heating and
solar wind acceleration.
Some simulation codes \citep[e.g.,][]{Sj04,Nita18} assign a constant
temperature and a hydrostatic density profile to all open-field regions.
\citet{Sj04} also discussed how the $L$-dependence of many proposed
heating models may go away for sufficiently long loops, and they
proposed a kind of saturation that caps the value of $L$ to be used
in coronal heating relations like those given above.
We use a similar approach, but for open field lines we simply replace
the traced-out value of $L$ (which ends up being truncated at the
source surface) by an effective value given by
\begin{equation}
  L_{\rm open} \, = \, \frac{\pi}{2} (R_{\rm ss} - R_{\odot})
  \label{eq:Lopen}
\end{equation}
where $R_{\odot}$ is the solar radius and $R_{\rm ss}$ is the
heliocentric radius of the source surface
(see Section~\ref{sec:plasma:mag}).
We then use Equations~(\ref{eq:poynt})--(\ref{eq:Qpractical}) 
for the heating rate with this effective loop length.
Our value for $L_{\rm open}$ is essentially the arc-length of a
quarter-circle with a radius that extends up to the source surface height.
We do not impose any $L$-saturation on the closed loops,
despite \citet{Sj04} suggesting that this kind of modification may need
to be imposed for loops longer than about 50~Mm.

Our assumption that the coronal heating behaves like
Equations~(\ref{eq:poynt})--(\ref{eq:Qpractical}) is a huge
simplification of an extremely complex reality.
In the future, we intend to explore more complicated scaling
relations, such as ones that depend on additional details about the
local magnetic topology or the rate of flux emergence
\citep[e.g.,][]{Pr18,Ag19,Lu24}.
We also do not yet implement the effects of ``Taylor relaxation''
in nonpotential active regions \citep{T74,vB08,Ya18}.
To compensate for the fact that our simulations do not yet account
for the full range of heating mechanisms in magnetically complex
active regions, the results in Section~\ref{sec:results:obs}
are listed separately for: (a) all data, and (b) only the data
outside active regions.

\subsection{Relations to Other Parameterizations}
\label{sec:heating:other}

Our Equation~(\ref{eq:Qpractical}) has been written in a way that
it can be directly compared to the coronal heating parameterization
used by \citet{Ma10},
\begin{equation}
  Q_{\rm heat} \,\, = \,\, H \, T^{\alpha} P^{\beta}
  \label{eq:heatMa10}
\end{equation}
where $H$ is a constant that depends linearly on the quantity in
square brackets in Equation~(\ref{eq:Qpractical}).
It may be a bit counterintuitive to parameterize the heating this way,
but the form appears to be flexible enough to describe many proposed
scenarios.
For example, the original constant heating-rate assumption of RTV
\citep[i.e.,][]{RTV} would correspond to $\alpha = 0$ and $\beta = 0$.
Values of $\alpha \neq 0$ and $\beta \neq 0$ provide a way to
parameterize heating rates that vary as a function of position $s$.
\citet{Ma10} assumed the basal pressure $P$ in
Equation~(\ref{eq:heatMa10}) to be a constant, and it can be coupled
to the ideal-gas equation of state,
\begin{equation}
  P \,\, = \,\, \frac{\rho k_{\rm B} T}{\langle m \rangle}
\end{equation}
where $k_{\rm B}$ is Boltzmann's constant and the mean atomic mass
$\langle m \rangle$ is often specified as 0.5 times the mass of a
hydrogen atom in a fully-ionized hydrogen plasma.
Matching the above to Equation~(\ref{eq:Qpractical}), we find
\begin{equation}
  \alpha \, = \, (m-2) \delta - \frac{m}{2}
  \,\,\, , \,\,\,\,\,\,\,
  \beta \, = \, \frac{m}{2} 
  \label{eq:alphabeta}
\end{equation}
and $H$ depends on both $m$ and $n$ as well.
We implement these exponents when solving the time-steady
equations of \citet{Ma10}; see Section~\ref{sec:plasma:Ma10}.

Comparisons between observations and models often describe the
heating rate in terms of primary dependences on magnetic field
strength and loop length,
\begin{equation}
  Q_{\rm heat} \, \propto \, B^a / L^b \,\,\, ,
\end{equation}
where we clearly distinguish the exponents $a$ and $b$ from
the ones defined above (i.e., $\alpha$ and $\beta$).
Some studies use $B_{\odot}$ in this relation and others use
$\langle B \rangle$, the field strength averaged over the loop.
Using $B_{\odot}$, \citet{Sj04} found best agreement with EUV and
X-ray data using $a=1$ and $b=2$.
On the other hand, using $\langle B \rangle$, \citet{Lu08} found
best agreement with data using $a=1$ and $b=1$.
\citet{WW06} found a connection between these results by determining
$\langle B \rangle \propto B_{\odot}/L$ for the active regions they studied.
If this were also true for the loops modeled by \citet{Sj04} and
\citet{Lu08}, then their results would be identical.
However, for different data, \citet{Wi08} found $a \approx 0.29$ and
$b \approx 1.24$, using the same mapping between
$\langle B \rangle$ and $B_{\odot}$, and \citet{DD11} found
$a \approx 0.7$--0.8 and $b \approx 0.5$.
In any case, if we match the above to
Equation~(\ref{eq:Qpractical}), we find
\begin{equation}
  a \, = \, 2 - m
  \,\,\, , \,\,\,\,\,\,\,
  b \, = \, 1 + n - m
  \label{eq:abdef}
\end{equation}
and the \citet{Sj04} case would correspond to $m=1$ and $n=2$.
This result, which appears to match the field-line braiding model 
of \citet{P83}, was also found by \citet{Is24}.
Note that Equation~(\ref{eq:abdef}) was used to produce some of the
ranges for $m$ and $n$ given for computer simulations at the bottom
of Table~\ref{table01}, since those simulation papers usually
provided $a$ and $b$ instead.

\section{Computing Time-Steady Plasma Properties}
\label{sec:plasma}

\subsection{The Magnetic Field}
\label{sec:plasma:mag}

To simulate the coronal heating at a specific date and time,
we extrapolate the Sun's magnetic field from synoptic (i.e.,
full-sphere) photospheric magnetograms.
For this purpose, we use the synoptic maps produced by the
Air Force Data Assimilative Photospheric Flux Transport (ADAPT)
project \citep{Arg10,Hk15},
with data taken from the National Solar Observatory's
Global Oscillation Network Group \citep[GONG;][]{Hv96,Hi18}.
The ADAPT maps combine data assimilation with photospheric flux
transport simulations to approach the goal of truly synchronic maps
of the Sun's full surface.
For additional discussion of the specific dates and times chosen for
this study, see Section~\ref{sec:emission}.

The field-line extrapolation is carried out using the standard
Potential Field Source Surface (PFSS) method, which assumes the
corona has zero current out to a spherical ``source surface''
at radius $R_{\rm ss} = 2.5 \, R_{\odot}$.
Above that height, the field is assumed to be oriented radially
\citep{Sh69,AN69}.
PFSS has been shown to produce relatively accurate mappings from the
Sun to the heliosphere \citep{AP00,Lu02,WS06}, although full MHD
simulations take better account of gas pressure effects and
solar-wind stream interactions.
When reconstructing the spherical harmonic coefficients from the
ADAPT magnetograms, we kept only terms in the expansion with degree
$\ell \leq 100$.
If there was any residual monopole field, we set it to zero.

For any given location in the corona, we need to know the length of
the loop that passes through that point and the strength of the
magnetic field at that loop's footpoint(s).
Thus, we perform numerical field-line tracing by computing the
local vector field, taking an explicit first-order Euler step in
that direction, recomputing the vector direction at the new location,
and repeating until we reach either the photosphere or the source
surface.
This process is then repeated, but in the direction anti-parallel
to the field.
Due to a decreasing level of magnetic complexity with increasing
height, we use a radius-dependent step size
\begin{equation}
  \Delta s \,\, = \,\,
  \mbox{(0.001 $R_{\odot}$)} \sqrt{\frac{r}{R_{\odot}}}
  \,\,\, .
\end{equation}
We integrate in both directions from an arbitrary start-point in the
corona to find the total loop-length, and we divide that by two to obtain
the half-length $L$.

\begin{figure*}[!t]
\epsscale{1.00}
\plotone{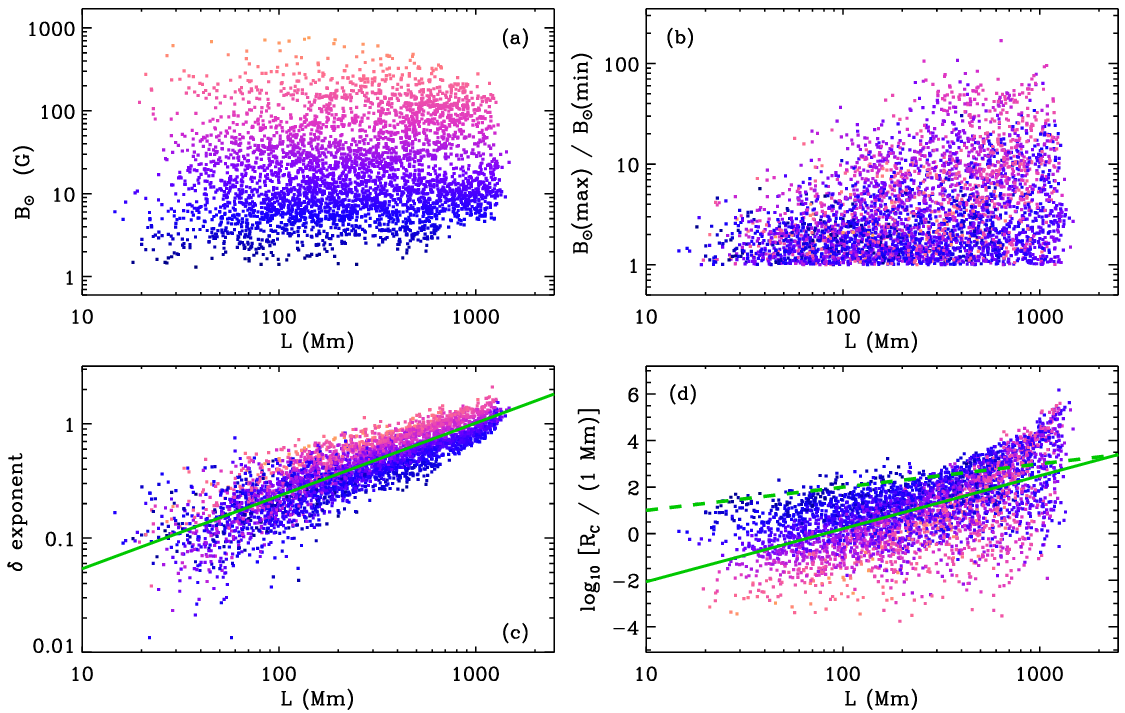}
\caption{Loop-length dependence of (a) the mean photospheric field
strength $B_{\odot}$, (b) the ratio of the photospheric field strength
at the stronger footpoint to that of the weaker footpoint, (c) an
estimate of the $\delta$ exponent from Equation~(\ref{eq:deltadef}),
(d) the local radius of curvature at each starting point.
Point-color corresponds to the value of $B_{\odot}$.
Also shown are best-fit power-law relationships given in the text
(solid green curves) and the one-to-one trend line for $L = R_{\rm c}$
(dashed green curve).
\label{fig01}}
\end{figure*}

The need to associate a value of $L$ with every point along each
line of sight (LOS) makes this the most time-consuming part of the
numerical calculation of coronal UV/X-ray emission.
In order to save computation time, others \citep[e.g.,][]{Mok16}
proposed replacing $L$ by a quantity proportional to the local radius
of curvature of the field lines,
\begin{equation}
  R_{\rm c} \,\, = \,\, \frac{B^2}{|{\bf B} \cdot \nabla {\bf B}|}
  \,\,\, .
\end{equation}
Below, we investigate to what extent $R_{\rm c}$ is a useful proxy for $L$.

As an illustrative example, we traced 5,000 field lines
sampled randomly through the volume of the corona extrapolated from
the ADAPT map of 2013 December 20, 18:00~UT.
For each starting point, we sampled its heliographic longitude
and the sine of its latitude from uniform random distributions.
We sampled the starting radii from a uniform distribution in the
logarithm of the height above the photospheric base (i.e., we sampled
the quantity $r/R_{\odot} - 1$ between minimum and maximum values of
0.008 and 1.49).
This was done to best mimic the distribution of starting points that
we expect to encounter when integrating through optically thin
lines of sight.
Figure~\ref{fig01} shows a selection of $L$-dependent quantities for
the closed loops identified by this process.
Note that about 78\% of the randomly chosen starting points intersected
closed loops and 22\% intersected open field lines that extend up
to the source surface.

All four panels of Figure~\ref{fig01} show that there does not seem
to be any dominant trend for loop occurrence as a function of $L$.
This result may appear to contradict other determinations of power-law
loop-length probability distributions extracted from data-driven
coronal field reconstructions \citep[e.g.,][]{Asch00,Asch08}.
In many of those studies, there appear to be many more short loops than
long loops.
However, those loops tend to be selected from distributions of
starting points on the solar surface.
On the other hand, our decision to select from a range of starting radii 
(between the photosphere and the source surface) must oversample the
longer loops.

Figure~\ref{fig01}(a) shows the mean photospheric field strength
$B_{\odot}$, which we define for closed loops as the geometric mean of
the absolute values of the photospheric field strengths at both ends.
Others \citep[e.g.,][]{DD11} use the arithmetic mean for their
symmetrized basal field strength.
To illustrate the range of asymmetry between the two footpoints,
Figure~\ref{fig01}(b) shows the ratio of the stronger to the weaker
photospheric field strengths.
The lack of large asymmetry ratios for the shortest loops may be
an artifact of the limited spatial resolution associated with
truncating the PFSS expansion at spherical harmonic index $\ell = 100$.
On the other hand, it may just be true that (for most of the coronal
volume) shorter loops are more symmetric than longer loops.
In any case, despite the presence of ratios up to about 150, the
distribution of ratios is very skewed.
For the full set of 5,000 points, the median value of the ratio is
only 2.5, so most loops are not extremely asymmetric in this sense.

Figure~\ref{fig01}(c) shows the $\delta$ exponent as defined in
Equation~(\ref{eq:deltadef}).
This quantity characterizes the overall decrease in field strength
from a loop's base to its apex.
For each traced loop, we kept track of the smallest value of the absolute
value of $B(s)$, and we assumed this location corresponds to a fiducial
loop-apex temperature of 3~MK.
Choosing different values of this temperature does not strongly
affect the resulting values of $\delta$.
There is a clear trend for longer loops to have larger values of
$\delta$, and the following power-law fit is also shown in
Figure~\ref{fig01}(c):
\begin{equation}
  \delta \, = \, \left( \frac{L}{\mbox{975 Mm}} \right)^{0.638}
  \,\,\, .
  \label{eq:delfit}
\end{equation}
Similar parameterizations of this kind were discussed by \citet{Sch16}.
This trend also implies that the ratio $\langle B \rangle / B_{\odot}$,
where $\langle B \rangle$ is the field strength averaged over the loop,
must also decrease as $L$ increases, as has been inferred from observations
\citep[e.g.,][]{Mn00,JM06,WW06} and is discussed in more
detail in Section~\ref{sec:results:disc}.

Lastly, Figure~\ref{fig01}(d) compares the radius of curvature
$R_{\rm c}$, computed at each loop's starting point, with the actual
loop-length $L$.
The distribution of $R_{\rm c}$ values is broad, but there does exist
a trend for longer loops to have larger radii of curvature.
The power-law fit to that trend is given by
\begin{equation}
  R_{\rm c} \, = \, \left( \frac{L}{\mbox{64 Mm}} \right)^{1.72}
\end{equation}
with the resulting values for $R_{\rm c}$ given in in units of Mm.
This curve falls somewhat below the one-to-one trend line consistent
with the assumption of $R_{\rm c} =  L$.
In fact, only 12\% of the plotted points fall within a multiplicative
factor of two of the $R_{\rm c} =  L$ curve.
Thus, we do not believe the local radius of curvature is a particularly
useful proxy for the loop length.

\subsection{Electron Thermal Conduction}
\label{sec:plasma:cond}

In the solar corona, heat conduction is an important process
that can transport energy over large distances and play a major role
in determining the spatial variation of temperature.
We use a general diffusive form for its scalar volumetric rate,
\begin{equation}
  Q_{\rm cond} \, = \, B \, \frac{\partial}{\partial s}
  \left( \frac{\kappa_e}{B} \, \frac{\partial T}{\partial s} \right)
  \,\,\, ,
  \label{eq:Qcond}
\end{equation}
where $\kappa_e$ is the heat conductivity due to free electrons.
For now, we assume the transport is dominated by electron heat
flux parallel to the background magnetic field.

In a fully ionized plasma, the simplest description of electron
conductivity is the \citet{SH53} model, which gives
$\kappa_e \propto T^{5/2}$ and depends on particle-particle
collisions being rapid.
However, there have been many studies of non-classical extensions
to the Spitzer--H\"{a}rm theory.
First, in low-density open-field regions far from the Sun,
particles stream freely over long distances between individual
collisions and the heat flux no longer depends on the
temperature gradient \citep[see, e.g.,][]{H74,CS21}.
Second, for exceedingly hot plasmas that vary rapidly on small
scales, the classical heat flux may saturate at an upper limit
described by pure advection at the thermal speed \citep{MK75,SL79}.
Third, in partially ionized plasmas neutral atoms can
drift relative to the ions and electrons, which results in
additional energy transport via ambipolar diffusion \citep{MS56}.

For the coronal loops to be modeled in this paper, we choose to
include only the last of the three non-classical modifications to
Spitzer--H\"{a}rm conductivity described above.
\citet{FAL90} produced semi-empirical models of the chromosphere,
transition region, and low corona that included the effects of
ambipolar diffusion.
Subsequently, \citet{KL99} and \citet{SvB05} analyzed those
models and explored low-temperature modifications to $\kappa_e$
that aimed to include this process in coronal heating models.
\citet{SvB05} proposed a temperature dependence of the form
\begin{equation}
  \kappa_e \, = \, \kappa_0 T^{5/2} \, + \, \kappa_1 T^{-1/2}
  \,\,\, ,
\end{equation}
where $\kappa_0$ was set to reproduce the classical \citet{SH53}
limit in the high-temperature limit and $\kappa_1$ was set so that
the minimum in the curve $\kappa_e(T)$ occurred at $10^5$~K.

\begin{figure}[!t]
\epsscale{1.19}
\plotone{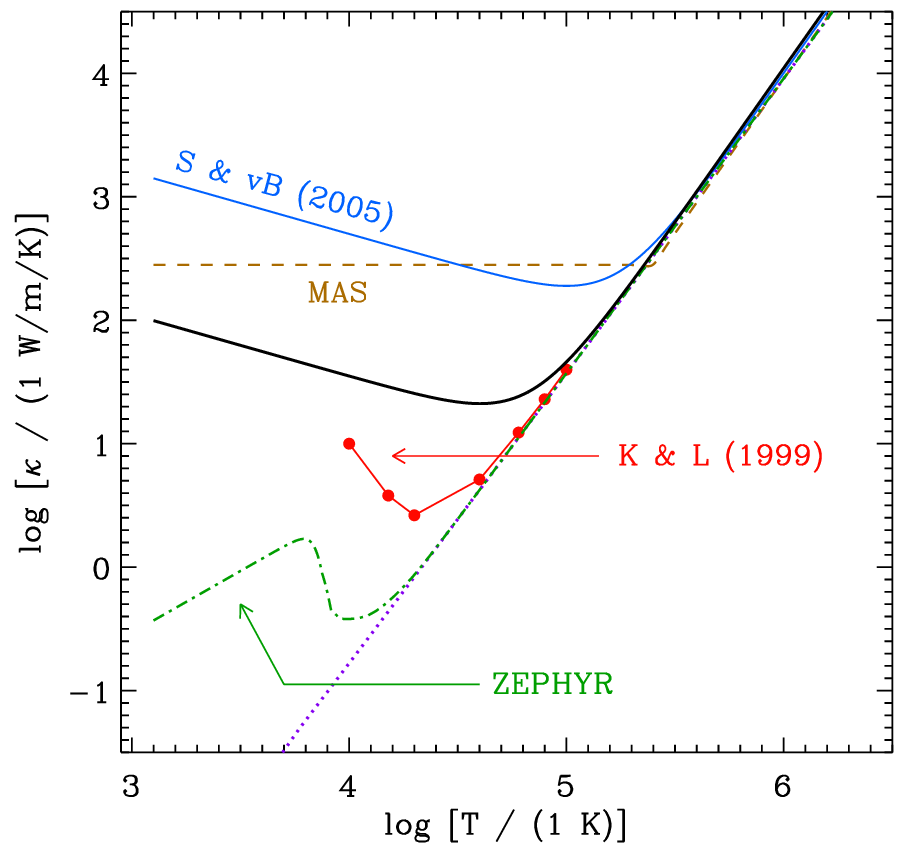}
\caption{Temperature dependences of electron heat
conductivity prescriptions:
classical \citep[][violet dotted line]{SH53}, and modified versions
from \citet{KL99} (red curve with points),
\citet{SvB05} (blue solid curve),
\citet{Lk01} (brown dashed curve),
and \citet{CvB07} (green dot-dashed curve).
Also shown is the form we adopt: Equation~(\ref{eq:kappamod})
with $T_{\rm A} = 40,000$~K (black solid curve).
\label{fig02}}
\end{figure}

Figure~\ref{fig02} illustrates the temperature dependence of this
modified conductivity and compares it with other estimates.
For example, \citet{KL99} found a slightly cooler temperature
for the minimum in $\kappa_e(T)$ than did \citet{SvB05}, as well
as a more rapid rise toward lower values of $T$.
The ZEPHYR model of \citet{CvB07} included partial ionization
effects for hydrogen that modified $\kappa_e$ at even lower
temperatures \citep[see also][]{NU77}.
Also, \citet{Lk01} implemented a similar low-temperature flattening 
in the Magnetohydrodynamics Around a Sphere (MAS) code as a
numerical  tool for resolving the transition region.
We show these curves in Figure~\ref{fig02} as well.

To move forward with the modeling,
we define an ambipolar transition temperature $T_{\rm A}$
as a free parameter and set up the modification as a multiplicative
correction factor,
\begin{equation}
  \kappa_e \, = \, \kappa_0 T^{5/2} \left[ 1 +
  5 \left( \frac{T_{\rm A}}{T} \right)^3 \right] \,\,\, ,
  \label{eq:kappamod}
\end{equation}
where the factor of 5 is necessary for the minimum in $\kappa_e(T)$
to occur exactly at $T_{\rm A}$.
In this paper, we use a typical Spitzer--H\"{a}rm value of
$\kappa_0 = 1.1 \times 10^{-6}$ erg~s$^{-1}$~cm$^{-1}$~K$^{-7/2}$.
This fixed value assumes that the Coulomb logarithm remains
roughly constant over the parameters of interest.
Also, we use $T_{\rm A} = 40,000$~K, as shown in Figure~\ref{fig02},
which is intermediate between the curves from \citet{KL99} and
\citet{SvB05}.

Curiously, \citet{Br19} found a similar temperature dependence
(i.e., $\kappa \propto T^{-1/2}$) for conduction dominated by
turbulent scattering in coronal loops.
However, they found this effect is most likely to dominate the
total conductivity in the hottest parts of a loop, whereas classical
collisional conduction would dominate in the coolest parts
\citep[see also][]{Dai24,AB25}.
This behavior is the opposite of the ambipolar correction given
in Equation~(\ref{eq:kappamod}), so we set it aside for now and
leave the study of other turbulent modifications to classical
heat conduction for future work \citep[see, e.g.,][]{HJ72,EB18,Hd22}.

Our use of a modified electron heat conductivity can be implemented as
a change of variables, from temperature $T$ to a dimensionless
variable $\eta$ that takes the details of heat conduction and the
varying magnetic field strength (i.e., $B \propto T^{-\delta}$)
into account.
This new variable is defined in order to satisfy
\begin{equation}
  Q_{\rm cond} T^{\delta} \, = \, \frac{\partial}{\partial s}
  \left( T^{\delta} \, \kappa_e \, \frac{\partial T}{\partial s} \right)
  \, \propto \, \frac{\partial^2 \eta}{\partial s^2} \,\,\, .
  \label{eq:etaneed}
\end{equation}
For classical heat conduction, \citet{Ma10} used
\begin{equation}
  \eta \, = \, \left( T / T_{\rm max} \right)^{3.5 + \delta}
  \label{eq:etaold}
\end{equation}
where the apex temperature $T_{\rm max}$ is discussed much more below.
Our new form for the electron conductivity is consistent with
a modified definition $\eta = C(T)/C(T_{\rm max})$, where
\begin{equation}
  C(T) \, = \, T^{\delta} \left[
  T^{3.5} + \left( \frac{10\delta + 35}{2\delta + 1} \right)
  T_{\rm A}^3 T^{1/2} \right] \,\, .
  \label{eq:etanew}
\end{equation}
This satisfies Equations~(\ref{eq:kappamod}) and (\ref{eq:etaneed}),
and it is used below for modeling the spatial distribution $T(s)$.

\subsection{Radiative Losses}
\label{sec:plasma:rad}

The rate of heat loss (or gain) due to photon emission (or absorption)
is denoted $Q_{\rm rad}$.
A complete description of this rate at all levels of the solar
atmosphere would require constructing a non-local thermodynamic
equilibrium (non-LTE) radiative transfer model consistent with the
solution for the plasma properties.
However, in this paper we are dealing only with regions above
the ``coronal base'' (i.e., $T \geq 10^4$~K).
Thus, we follow a long history of using the limiting case of an
{\em optically thin} radiative-loss rate
\begin{equation}
  Q_{\rm rad} \, = \, -n_{e}^{2} \, \Lambda(T)
  \label{eq:Qrad0}
\end{equation}
where $n_e$ is the electron number density and $\Lambda(T)$ is
a tabulated radiative-loss function (sometimes called the
cooling curve) that takes account of both continuum and line
opacities \citep{CT69,AA89,Ma00}.
In some other work, the squared density factor above is given as
$n_e n_{\rm H}$, where $n_{\rm H}$ is the total hydrogen number
density.
We choose the version in Equation~(\ref{eq:Qrad0}) that presumes
full hydrogen ionization and also includes a model of coronal
ionization equilibrium for all elements.

\begin{figure}[!t]
\epsscale{1.20}
\plotone{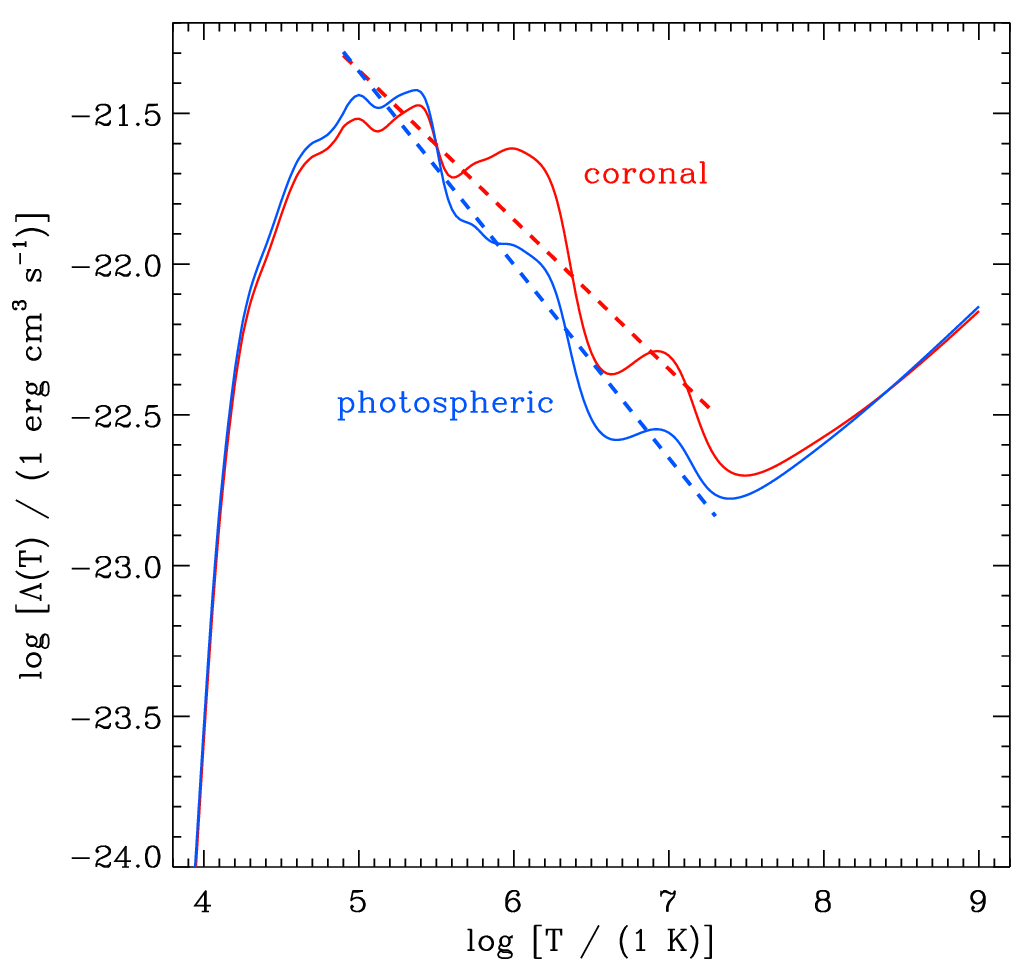}
\caption{Optically thin radiative-loss functions for photospheric
(blue curves) and coronal (red curves) abundances.
Tabulated data (solid curves) are compared with best-fitting
power-law fits (dashed curves).
\label{fig03}}
\end{figure}

We take this opportunity to describe a new form of $\Lambda(T)$ that
consists of three components stitched together:
\begin{enumerate}
\item
For $T < 10^{4.05}$~K, the adopted functional form was given by
the analytic non-LTE model of \citet{GJ12} \citep[see also][]{KP91}.
We modify this by including a constant lower limit
$\Lambda_{\rm min} = 10^{-34}$ erg~cm$^3$~s$^{-1}$ that accounts
for a photoionization ``floor'' at low temperatures; something like
this has been used for stars cooler than the Sun \citep{Sch03}.
\item
For $10^{4.05} < T < 10^{4.9}$~K, we used the same optically thin
radiative-loss function as \citet{CvB07}.
This was based on a semi-empirical non-LTE chromosphere model
that was thereafter published by \citet{AL08}.
A key feature in this part of the curve is the lack of a prominent
local maximum, due to H~I Ly$\alpha$, that appears at
$T \sim 10^{4.1}$~K in versions of $\Lambda(T)$ computed without
reference to a specific background atmosphere
\citep[see also][]{Kl08,CL12}.
\item
For $T > 10^{4.9}$~K, we used the radiative-loss function computed
by version 10 of the CHIANTI atomic database \citep{De97,DZ21}.
To account for very weak dependences on density, this part was computed
as the geometric mean of two models with fixed electron densities
of $10^{7}$ and $10^{15}$~cm$^{-3}$.
The two original $\Lambda$ curves never differed from one another
by more than about 20\%, and the weak density dependence will be
studied in future applications of this curve to models of solar wind
acceleration.
\end{enumerate}
Figure~\ref{fig03} shows the adopted radiative-loss function for two
sets of element abundances used in the CHIANTI calculation:
photospheric \citep{Asp09,Grv15,Sco15} and coronal \citep{Schm12}.
The latter set includes the effects of measured abundance enhancements
for some elements with low first ionization potential (FIP) values.
In Section~\ref{sec:params:1st} we discuss the results of
tests that explore the sensitivity of coronal properties to the choice
of cooling curve.

Figure~\ref{fig03} also displays the result of power-law fits to
the parts of these $\Lambda(T)$ functions that are most relevant
for computing the time-steady properties of coronal loops.
These kinds of power-law fits have been used to enable analytic
solutions to the coronal thermal energy equilibrium equations
\citep[see, e.g.,][]{RTV,KM82,Ma00} and we continue that use
in this paper.
For an assumed constant gas pressure given by $P_0 = 2 n_e k_{\rm B} T$,
the power-law behavior can be written as
\begin{equation}
  Q_{\rm rad} \, = \, - P_0^2 \, \chi_0 \, T^{-2-\gamma} \,\,\, ,
  \label{eq:radMa10}
\end{equation}
where $\gamma$ and $\chi_0$ are the fit parameters describing
$\Lambda(T)$ and the other physical constants in this expression.
We used tabulated data for $\log T$ between 4.9 and 7.3,
similar to the range used by \citet{Ma00}, and performed
least-squares fits to straight lines in log-log space.
For photospheric abundances, the resulting parameters were
$\gamma = 0.6413$ and $\chi_0 = 9.198 \times 10^{12}$,
with the latter assuming cgs units for all quantitites in
Equation~(\ref{eq:radMa10}).
For coronal abundances, the best-fit parameters were found to be
$\gamma = 0.4955$ and $\chi_0 = 1.729 \times 10^{12}$.
Also, for reference, the constants used by \citet{Ma10} were
$\gamma = 0.5$ and $\chi_0 = 2.570 \times 10^{12}$, intermediate
between our two sets of values.

\subsection{Solving the Steady-State Energy Equation}
\label{sec:plasma:Ma10}

Our goal is to compute local values of the temperature and density,
$T(s)$ and $\rho(s)$, each as a function of spatial position $s$
along any given loop.
Following \citet{Ma10}, we convert the time-steady energy conservation
equation into a nondimensional form by using a spatial coordinate
$x = s/L$ and the temperature variable $\eta$ defined above.
The new form for $\eta$, defined by Equation~(\ref{eq:etanew}),
is used for $Q_{\rm cond}$, but we continue to assume the old form,
Equation~(\ref{eq:etaold}), when writing the nondimensional terms
corresponding to $Q_{\rm heat}$ and $Q_{\rm rad}$.
This inconsistency is unimportant over most coronal and
transition-region temperatures (i.e., $T \gtrsim 10^5$~K), but we
will investigate more self-consistent approaches in future work
\citep[see also][]{Dai24,AB25}.

With the above caveats, Equations (\ref{eq:Qbal}), (\ref{eq:heatMa10}),
(\ref{eq:etaneed}), and (\ref{eq:radMa10}) can be written as
\begin{equation}
  \epsilon \, \frac{d^2 \eta}{dx^2} \,\, = \,\,
  \eta^{\mu} - \xi \eta^{\nu}
\end{equation}
where the exponents are known,
\begin{equation}
  \mu \, = \, \frac{\delta - 2 - \gamma}{3.5 + \delta}
  \,\, , \,\,\,\,\,
  \nu \, = \, \frac{\alpha + \delta}{3.5 + \delta}
  \label{eq:munu}
\end{equation}
and the coefficients involve unspecified parameters $P_0$ and
$T_{\rm max}$,
\begin{equation}
  \epsilon \, = \, \frac{\kappa_0 T_{\rm max}^{5.5 + \gamma}}
  {(3.5 + \delta) L^2 P_0^2 \chi_0}
  \label{eq:epsdef}
\end{equation}
\begin{equation}
  \xi \, = \, \frac{H T_{\rm max}^{\alpha + \gamma + 2}}
  {P_0^{2-\beta} \chi_0} \,\, .
  \label{eq:xidef}
\end{equation}
By applying four independent boundary conditions to the second-order
differential equation given above, \citet{Ma81,Ma10} discussed how to
specify the $\epsilon$ and $\xi$ parameters in terms of known quantities,
\begin{equation}
  \xi \, = \, \frac{\nu + 1}{\mu + 1}
  \,\, , \,\,\,\,\,
  \epsilon \, = \, \frac{2 (\nu - \mu)^2}
  {(\mu + 1) [B(\lambda + 1, 0.5)]^2}
  \label{eq:exknown}
\end{equation}
where $B(x,y)$ is the complete beta function and
$\lambda = (1 - 2\nu + \mu)/(2\nu - 2\mu)$.
In the limit of $\alpha = \beta = \delta = 0$,
Equations (\ref{eq:munu})--(\ref{eq:exknown}) essentially give
the two ``RTV scaling laws'' of \citet{RTV}.
Note that $B(\lambda + 1, 0.5)$ is well-defined only for
$\lambda + 1 > 0$, which corresponds to $\alpha > -(2+\gamma)$.
Models that violate this inequality are susceptible to thermal
non-equilibrium \citep[see also][]{KM82}.
In practice, if a given loop's value of $\alpha$ is computed
from Equation~(\ref{eq:alphabeta}) to be less than --2.25,
we truncate it at this effective floor value.

With the $\xi$ and $\epsilon$ values fixed by Equation~(\ref{eq:exknown})
and the heating-rate constant $H$ specified by the quantities given
in Section~\ref{sec:heating}, values for the undetermined parameters
$T_{\rm max}$ and $P_0$ can be found from
Equations~(\ref{eq:epsdef})--(\ref{eq:xidef}).
Of course, in order to specify $H$ we must have a value for the
normalization parameter ${\cal E}_0$, which has not yet been discussed.
Our general approach to specifying this parameter follows on from,
e.g., \citet{Sj04} and \citet{WW06}, who defined a ``typical'' coronal
loop with known $L$, $B_{\odot}$, and $T_{\rm max}$.
The normalization for the heating rate that maintains such a loop
is then used for other loops.
Thus, our typical loop is specified as one with $L = 15$~Mm,
$B_{\odot} = 100$~G, and $\delta$ given by Equation~(\ref{eq:delfit}).
Its fixed apex temperature is called $T_{\rm norm}$, and for a given
choice of coronal heating parameters (e.g., $m$ and $n$)
the above equations are used to solve for $H$ and $P_0$, and then for
${\cal E}_0$.
That heating-rate normalization is then applied to all other loops
encountered by lines of sight in this particular model.
Lastly, if we are searching for optimal agreement with observational
data, we vary $T_{\rm norm}$ as a free parameter together with
varying the $m$ and $n$ exponents.

\subsection{Spatial Dependence of Loop Properties}
\label{sec:plasma:spatial}

The equations described in Section~\ref{sec:plasma:Ma10} provide
solutions for the constants $T_{\rm max}$ and $P_0$, and these anchor
the normalizations of spatially dependent solutions for $T(s)$, $P(s)$,
and $n_e (s)$.
Starting with the temperature, we use the procedure described in
Section~3.5 of \citet{Ma10}.
This yields a solution for the dimensionless variable $\eta (x)$ that
involves the incomplete beta function.
Then, we use Equation~(\ref{eq:etanew}) when converting from $\eta(x)$
to the spatially dependent temperature $T(s)$.
Although the constants $T_{\rm max}$ and $P_0$ do not depend on
the chosen value of the ambipolar diffusion temperature $T_{\rm A}$,
the spatial dependence of $T(s)$ does depend on it.
We ensure the chromospheric lower boundary condition ($\eta = 0$ at $x=0$)
is scaled to our adopted base temperature
$T_{\rm base} = 10^4$~K.

Because a hot corona exhibits large scale heights, a reasonable first
guess for the gas pressure has traditionally been an {\em isobaric}
atmosphere with $P(s) \approx P_0$.
In fact, the equations given in Section~\ref{sec:plasma:Ma10}
were solved under this isobaric assumption.
However, when forward-modeling the plasma properties of large coronal
loops, it is helpful to attempt to improve upon this simple model.
Modifications that account for $P(s) \neq P_0$ come in two distinct types:
(1) solving more involved equations for non-isobaric values of the
constants $T_{\rm max}$ and $P_0$, and
(2) accepting the earlier values of $T_{\rm max}$ and $P_0$,
but implementing a self-consistent hydrostatic spatial drop-off in
the ratio $P(s)/P_0$.
For now, we neglect the first type of modification because it tends to be
small in magnitude for coronal loops that dominate the EUV and X-ray
emission \citep[see, e.g.,][]{Se81,AS02,DD09}.
The second type of modification is discussed below.

We implement an equation of hydrostatic equilibrium that assumes a
semicircular shape for the coronal loop and accounts for the variation
in the direction of gravity at different values of $s$
\citep[see also][]{AS02}.
Specifically, for a semicircular loop with half-length $L$, its
circular radius is given by $2L/\pi$, and we define an angle
$\theta = \pi s / 2L$ that increases from 0 (at the base) to
$\pi/2$ (at the apex).
We also account for non-constant gravity above a spherical Sun with
radius $R_{\odot}$, so that the magnitude of the gravitational
acceleration at any point along the loop is given by
\begin{equation}
  |g| \, = \, g_0 \left( 1 + \frac{2L}{\pi R_{\odot}} \sin\theta
  \right)^{-2} \,\,\, ,
\end{equation}
where $g_0 = GM_{\odot}/R_{\odot}^2$.
The associated equation of hydrostatic equilibrium is
\begin{equation}
  \frac{dP}{ds} \, = \, -\rho |g| \cos\theta \,\,\, .
  \label{eq:hydrostatic}
\end{equation}
If we also assume an isothermal loop with $T(s) \approx T_{\rm max}$,
the integration of Equation~(\ref{eq:hydrostatic}) is straightforward,
and we obtain
\begin{displaymath}
  P(s) \, = \, P_0 \, \exp \Bigg\{ -\frac{2L}{\pi h_{\rm max}}
  \,\,\, \times
\end{displaymath}
\begin{equation}
  \times \,\,\,
  \sin \left( \frac{\pi s}{2L} \right)
  \left[ 1 + \frac{2L}{\pi R_{\odot}}
  \sin \left( \frac{\pi s}{2L} \right) \right]^{-1} \Bigg\}
  \label{eq:Psiso}
\end{equation}
where $h_{\rm max}$ is the isothermal scale height associated with
$T_{\rm max}$.
Once both $P(s)$ and $T(s)$ are known, we use the fully-ionized
equation of state to solve for the spatial variation in
$n_e = P / (2 k_{\rm B} T)$.

Our expressions for $T(s)$, $P(s)$, and $n_e (s)$ depend primarily
on the dimensionless spatial coordinate $x = s/L$.
However, the exact way to specify the length coordinate $s$ for an
arbitrary location along a loop has not yet been described.
According to the procedure described in Section~\ref{sec:plasma:mag},
we start at a known point in the corona (at which we need to evaluate
the plasma parameters) and we trace along the field in both directions.
For closed loops, we can define $L_1$ and $L_2$ as the lengths traced
toward both photospheric footpoints.
For simplicity, we define $s$ as the minimum of $L_1$ and $L_2$,
effectively ensuring that it is never larger than $L = (L_1 + L_2)/2$.
For open field lines, $s$ is the traced distance along the one direction
that ends up back at the photosphere.

\subsection{Time-Dependent Effects}
\label{sec:plasma:DEM}

There is abundant evidence for time variability in the corona,
either from large numbers of intermittent energy-release events
\citep[e.g.,][]{Lin84,EV99,VK11,Ir15,Bd21,Tw23}
or from thermal nonequilibrium cycles that can occur naturally even
for some forms of steady heating \citep{Mik13,Do16,Sc24}.
Because the timescales of heating and cooling are finite and generally
unequal to one another, there must be observable signatures of the
variability in the time-averaged properties of the plasma.
For example, \citet{Gu13} found that the precise shape of the cool
side of the mean differential emission measure (DEM) distribution
depends on the frequency of nanoflare-like heating events.
Much additional exploration of this behavior has been undertaken with
``zero-dimensional'' models that simulate a given loop's spatially
averaged time-dependence \citep[see, e.g.,][]{Kl08,Cg12,Chit13,Mond23}.

\begin{figure}[!t]
\epsscale{1.19}
\plotone{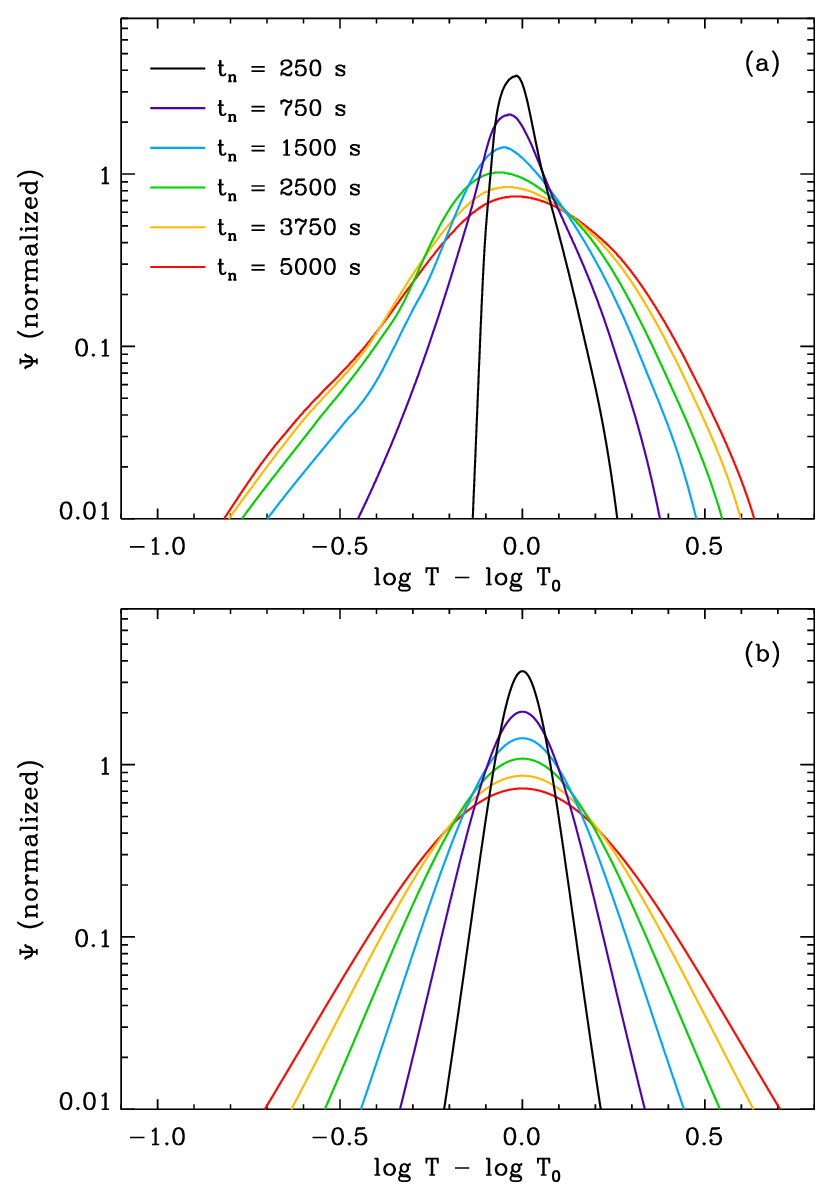}
\caption{Emission measure distributions from:
(a) time-dependent nanoflare simulations of \citet{Ba16} that
exhibited $T_0 = 4$~MK, and (b) parameterized fits described by
Equation~(\ref{eq:Psifit}).
Curves of different colors denote models with different
waiting times between nanoflare events ($t_n$); see inset for key.
\label{fig04}}
\end{figure}

One primary signature of underlying time dependence in $Q_{\rm heat}$
is the width of the resulting DEM as a function of temperature.
Very rapid nanoflaring ought to result in a narrow range of
coronal temperatures, whereas slow nanoflaring (with long intervals
allowing for substantial cooling and conductive transport) results
in a broad range of temperatures.
In this paper, we make use of time-averaged DEMs computed from the
stochastic nanoflare simulations made by \citet{Ba16}.
Figure~\ref{fig04}(a) shows the emission measure distributions from
their Figure~3; specifically the subset of models that assumed the
magnitude of each energy-release event scales linearly with the time
since the last event.
These models are parameterized by the mean waiting time $t_n$, which
was varied between 250 and 5000~s.
As expected, larger values of $t_n$ correspond to broader temperature
distributions.

In Figure~\ref{fig04}(b), we show the result of an analytic fit
whose parameters were adjusted iteratively in order to match various
aspects of the \citet{Ba16} curves.
We refer to this emission measure distribution as $\Psi$ and
define the fitting function as
\begin{equation}
  \Psi (T,T_0) \, \propto \, \left\{ 1 + \frac{1}{\kappa \sigma^2}
  \left[ \log_{10} \left( \frac{T}{T_0} \right) \right]^2
  \right\}^{-\kappa-1}
  \label{eq:Psifit}
\end{equation}
\cite[see, e.g.,][]{Ob68,Vs68},
and the normalization is computed numerically so that
\begin{equation}
  \int_0^{\infty} dT \,\, \Psi (T,T_0) \,\, = \,\, 1 \,\,\, .
\end{equation}
In the above expression, we use $\kappa = 4$ and 
\begin{equation}
  \sigma \, = \, 0.14 \left( \frac{t_n}{\mbox{1000 s}} \right)^{0.48}
  \,\, .
  \label{eq:sigmadef}
\end{equation}
Note that in the limit of $t_n \rightarrow 0$, we essentially assume that
$\Psi(T,T_0)$ reduces to a Dirac delta function centered on $T = T_0$.
By including the mean temperature $T_0$ as a free parameter, we also
assume that these curves can be shifted arbitrarily to different
central temperatures.
However, this is an extrapolation that goes beyond the simulation
results presented by \citet{Ba16}.

Our parameterized distributions are used to simulate the EUV and X-ray
emission from the corona.
The heating model described above predicts a unique value of $T(s)$ at
an arbitrary point in the coronal volume.
However, if $t_n$ is chosen to be finite, we then assume that point
contains unresolved ``multithermal'' plasma with a statistical distribution
of temperatures surrounding the central modeled value $T_0$.
Section~\ref{sec:emission:sim} describes how this leads to the
treatment of $\Psi (T, T_0)$ as a sliding convolution kernel for the
optically thin temperature response functions.

\section{EUV and X-ray Emission}
\label{sec:emission}

\subsection{Observational Data}
\label{sec:emission:obs}

We make use of high-resolution EUV narrow-band images from the
Atmospheric Imaging Assembly \citep[AIA;][]{Lm12} on the
Solar Dynamics Observatory \citep[SDO;][]{Ps12}, which was
launched in 2010.
AIA images have 4096$\times$4096 pixels, with typical
angular sizes of 0.6$''$ per pixel, and are routinely taken
with a cadence of 12~s in seven EUV bands.
We use AIA synoptic data, which are Level-1.5 binned
(i.e., 1024$\times$1024 pixel) images processed at a two-minute
cadence through the standard calibration routines \citep{Boe12,Boe14}
and are put online at the Joint Science Operations Center (JSOC) at
Stanford University.
These calibration steps include corrections for the detector flat-field,
dark-count subtraction, removal of anomalous data spikes, and
positional alignment to heliographic north.
We subsequently corrected for instrument-scattered light, which can
be a major source of contamination in dimmer regions such as coronal
holes, using the deconvolution routines provided by \citet{Hf25}.

We also use imaging data from the X-Ray Telescope \citep[XRT;][]{Go07}
on board the Hinode spacecraft \citep{Ko07}, which was launched in 2006.
XRT images have 2048$\times$2048 pixels, with typical angular sizes
of 1$''$ per pixel.
We use Level-2 data from the XRT synoptic program \citep{Tk16} which
takes full-Sun sets of images in multiple filters several times per day
and bins the data to 1024$\times$1024 pixel resolution.
These data have been processed through the standard calibration routines
(e.g., dark-count subtraction and intensity normalization), known
contamination spots on the detector have been removed, and satellite
pointing information has been included \citep[see also][]{Nk11,Nk14}.
We also corrected XRT data for instrument-scattered light using the
point-spread function discussed by \citet{Af16} and included as a
deconvolution in the {\tt XRTpy} package by \citet{Vq24}.
We ended up commenting out one line of the code that imposes an
intensity saturation above a data number (DN) of 2500.
That step appears to be required for flare-saturated
pixels, but our data did not include any flare events.

For comparison between observational data and the forward-modeling
process discussed above, we chose two time periods: one at the minimum
of the Sun's activity cycle and one at the maximum.
Also, for ease of modeling, we chose times when the heliographic
latitude of the Earth (often denoted $B_0$) was closest to zero.
This allowed our simulated lines of sight to be perpendicular to
the Sun's rotation axis.
Lastly, after searching for times that satisfied the above criteria,
we also optimized for the largest number of available XRT filters
in use during times when AIA was observing.
Thus, for the solar maximum data, we selected
2013 December 20, with XRT images obtained between 17:42 and 17:45 UT,
AIA images obtained at 17:44 UT, and the ADAPT/GONG magnetograms
from 20:00 UT.
For the solar minimum data, we chose 2018 December 7, with
XRT images obtained between 18:17 and 18:21 UT, AIA images
obtained at 18:18 UT, and ADAPT/GONG magnetograms from 18:00 UT.
In the remainder of this paper, we refer to these two data-sets
as 2013 (solar maximum) and 2018 (solar minimum).

\begin{figure*}[!t]
\epsscale{0.95}
\plotone{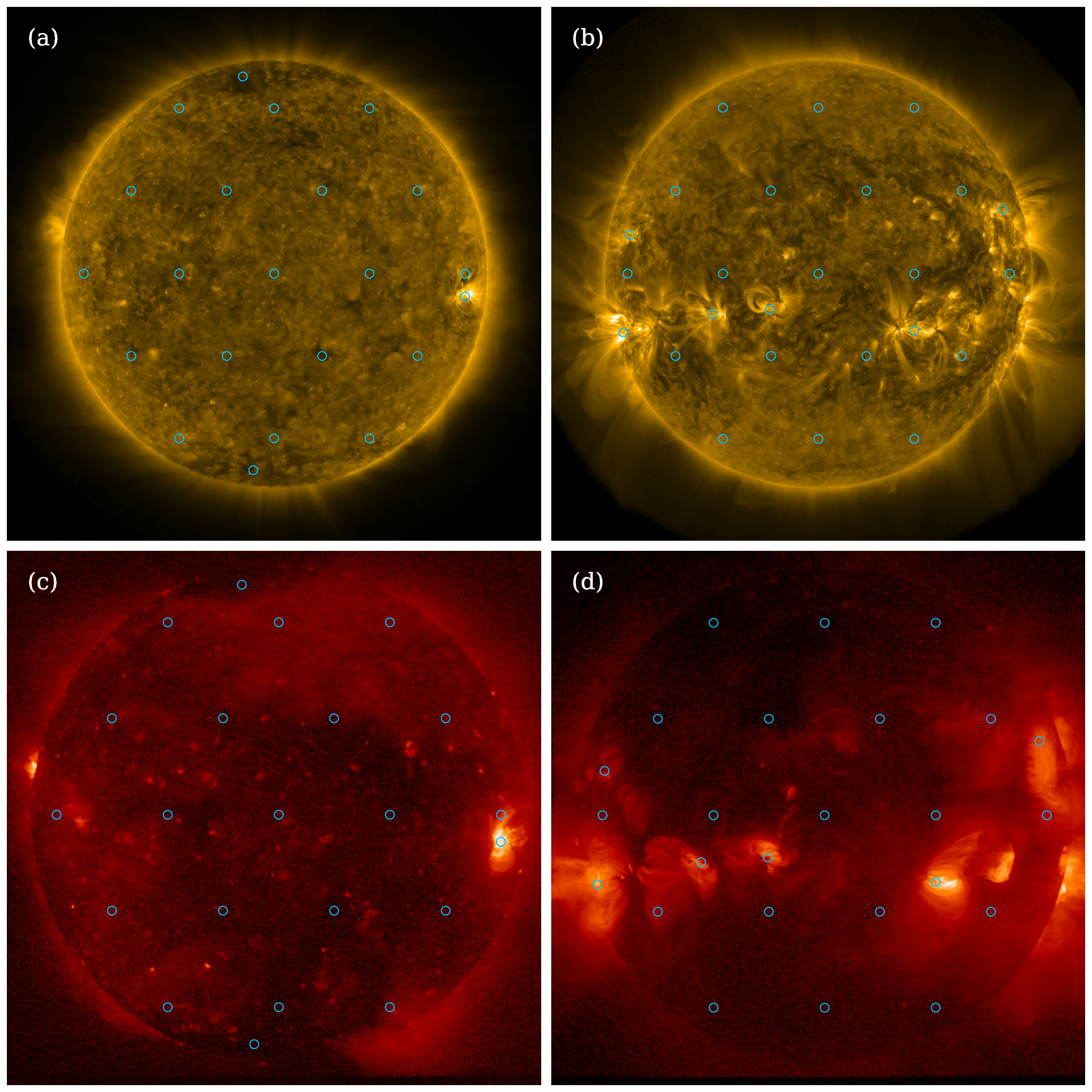}
\caption{(a) AIA 17.1~nm image at solar minimum (2018),
(b) AIA 17.1~nm image at solar maximum (2013),
(c) XRT Be-thin image at solar minimum (2018),
(d) XRT Be-thin image at solar maximum (2013).
Cyan circles show the locations of the LOS rays chosen
for model-based intensity synthesis.
\label{fig05}}
\end{figure*}

Figure~\ref{fig05} shows example data at these two times,
superimposed with locations on the solar disk intercepted by
our chosen sets of LOS points.
Each point is the base of a linear ``observing ray'' over which
we must integrate the EUV and X-ray emissivity (see below).
In this paper we used only LOS rays that intercept the solar disk,
and avoided off-limb rays, because the coronal heating model
for open field lines is still somewhat simplistic and does not
take account of solar wind acceleration.
For both time periods, we started with a grid of 19 LOS points
arrayed in a hexagonal pattern; i.e., packed equilateral triangles
with edge-lengths of 0.45~$R_{\odot}$.
Then, we positioned a number of additional rays to ensure that
some specific features were included in the modeling.
At solar minimum, we added only three more points---one for an
on-disk active region and two for the northern and southern polar
coronal holes---for a total of 22 rays.
At solar maximum, we added six more points---all in active
regions---for a total of 25 rays.

\subsection{Synthesizing Simulated Data}
\label{sec:emission:sim}

\begin{figure}[!t]
\epsscale{1.19}
\plotone{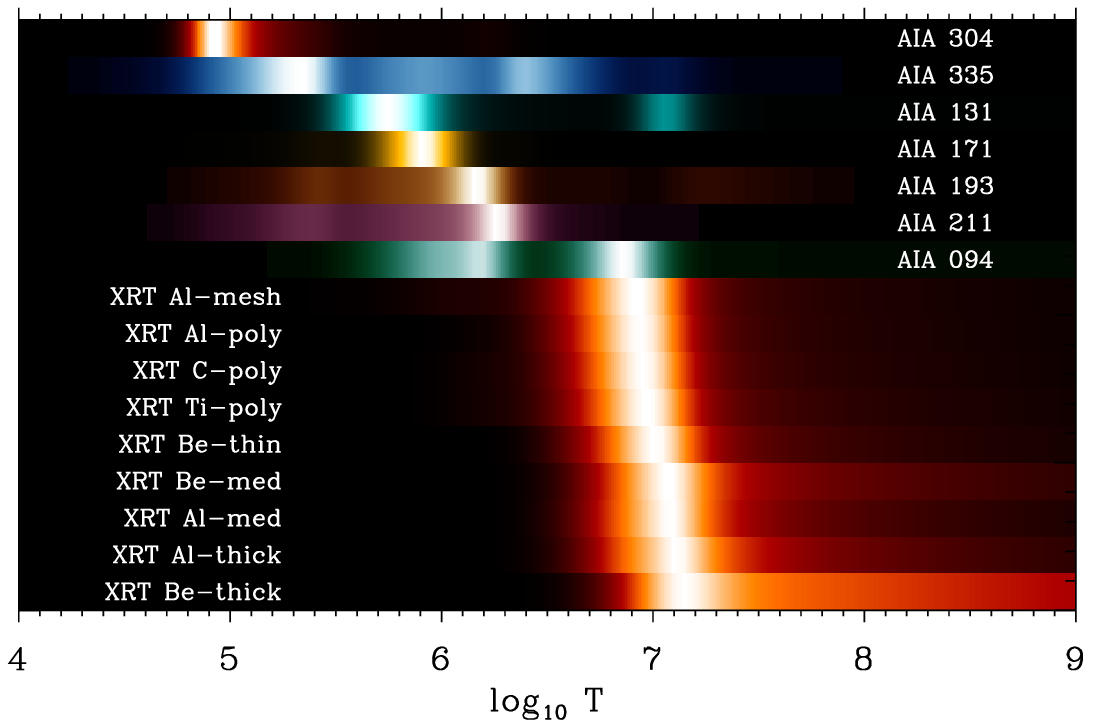}
\caption{Illustration of temperature response functions for
AIA and XRT, shown as brightness using the standardized
color maps for images in different filters and plotted as a function
of the base-10 logarithm of temperature in K.
Instrumental time degradation has been taken into account to produce
these functions for our solar maximum time period (2013 December 20).
\label{fig06}}
\end{figure}

When simulating the AIA and XRT emission along a given LOS ray,
we assume pure collisional excitation (i.e., electron impact
excitation) followed by spontaneous radiative de-excitation.
Thus, we neglect photon scattering as a potential source of emission
and also assume collisional ionization equilibrium.
For a given filter $i$, we compute the intensity $I_i$
(in units of DN s$^{-1}$ pixel$^{-1}$) as
\begin{equation}
  I_i \, = \, \int dx \,\, j_i (x)
  \, = \, \int dx \,\, n_e^2 \,\, K_i (T)
  \label{eq:intensity}
\end{equation}
where $x$ is the spatial coordinate along the LOS,
$j_i (x)$ is the spatially dependent emissivity at a given location,
and $K_i (T)$ is a given filter's temperature response function in
units of DN s$^{-1}$ pixel$^{-1}$ cm$^5$.
We use a Cartesian coordinate system with its origin
at the center of the Sun, the $x$-axis pointing toward the observer
(i.e., all observing rays are assumed to be parallel),
the $y$-axis pointing to solar west (i.e., to the right in standard
solar images), and the $z$-axis pointing to solar north.

The EUV and X-ray response functions are computed as integrals over
wavelength of the product of a filter effective area and a modeled
plasma emissivity at a specified temperature
\citep[see, e.g.,][]{Nk11,DZ13,Boe14}.
For AIA, we used version 10 of the routine {\tt aia\_get\_response}
from the IDL SolarSoft package \citep{FH98,FB00}, and this uses the CHIANTI
database \citep{De97,DZ21} to compute the required emissivity model.
For XRT, we used version 0.5.0 of the {\tt XRTpy} package \citep{Vq24}
to compute response functions, and this employed a CHIANTI emissivity
model constructed with the coronal abundances of \citet{Fd92}.

Figure~\ref{fig06} provides an illustrative summary of the shapes
of the seven EUV filters from AIA and the nine X-ray filters from XRT.
Note that the XRT temperature response functions are denoted by one of
the two instrument filter positions; in each case the other filter is
assumed to be in the open position.
For both the AIA and XRT detectors, we included known time degradation
effects by creating a grid of response functions between the years 2010
and 2023, then interpolating to the desired solar minimum and maximum dates.

For the comparisons between models and observations performed in this
paper, we neglected the AIA 30.4 nm filter (which primarily samples the
upper chromosphere) as well as the Al-thick, Be-med, and Be-thick
filters of XRT (which mostly sample hot flare plasma and typically have
negligibly low count rates over most of the solar disk).
In addition, for the 2018 time period we also could not use the
C-poly or Ti-poly filters, which had become contaminated and are
not recommended for use after 2015 June 14 \citep{Tk16}.
When combining the EUV and X-ray data, we ended up using 12 filters
in 2013 and 10 filters in 2018.

\begin{figure}[!t]
\epsscale{1.19}
\plotone{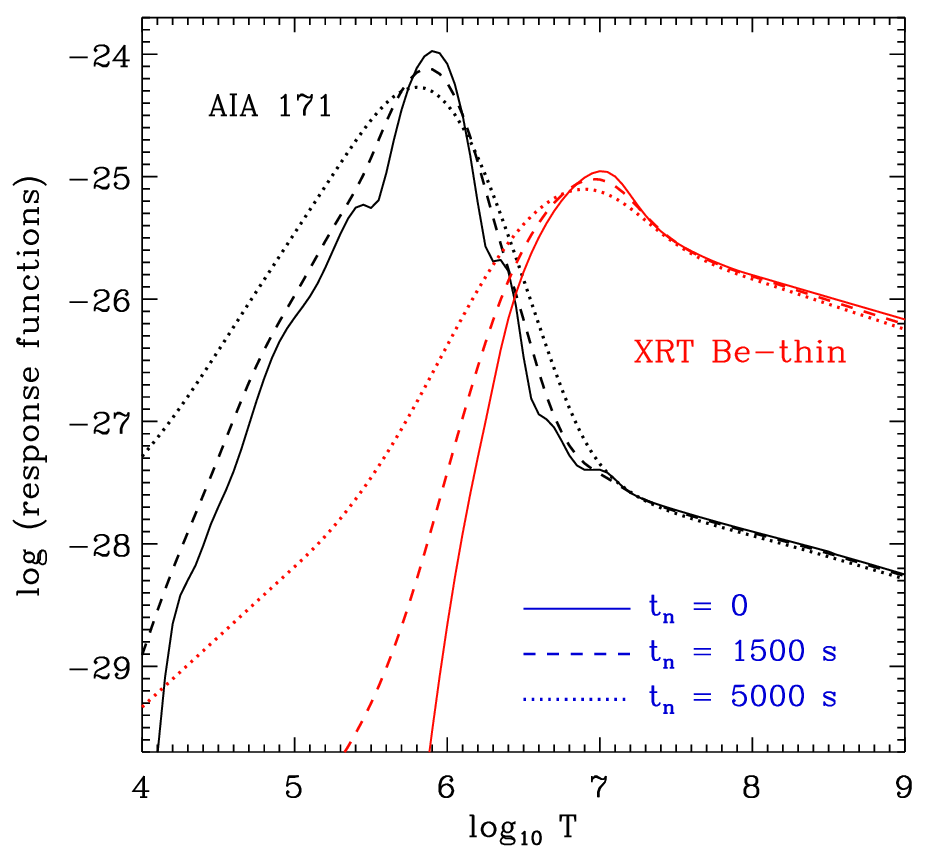}
\caption{Nanoflare-broadened temperature response functions for the
AIA 17.1~nm filter (black curves) and XRT Be-thin filter (red curves).
Values of $t_n = 0$ (solid curves), 1500~s (dashed curves), and
5000~s (dotted curves) are shown.
\label{fig07}}
\end{figure}

As discussed in Section~\ref{sec:plasma:DEM}, we took account of the
time-averaged multithermal behavior of a nanoflaring corona by
convolving the instrumental temperature response functions ($K_i$)
with the sliding broadening kernels ($\Psi$) defined above.
Thus, for models in which we assume nonzero values of the
mean nanoflare waiting time $t_n$, the modified temperature response
functions are given by
\begin{equation}
  \tilde{K}_{i} (T, t_n) \,\, = \,\,
  \int_0^{\infty} dT' \,\, K_i (T') \, \Psi (T', T) \,\,\, .
\end{equation}
Figure~\ref{fig07} shows the result of this kernel convolution
for two example filters (AIA 17.1~nm and XRT Be-thin) and
several choices of $t_n$.
It is worth noting that these response functions ignore some effects
that may be important to include when constructing more accurate
coronal loop models.
For example, including bulk plasma flows along the loops and
non-equilibrium ionization would certainly result in changes to
the adopted forms for $\tilde{K}_i(T,t_n)$.
In addition, taking account of the presence of suprathermal
electrons---especially for
regions in temporal and spatial proximity to flaring events---may
broaden the response functions in similar ways as shown in
Figure~\ref{fig07} \citep[see, e.g.,][]{Dz15}.

\section{Parameter Studies and Initial Results}
\label{sec:params}

Prior to presenting detailed comparisons between coronal heating
models and observational data, we describe a few initial results
of the modeling process described above.
These results tend to involve simplified assumptions or
background parameters, and they are meant to illustrate the
overall behavior of the coronal heating models.

\subsection{Idealized Grid of Coronal Loops}
\label{sec:params:1st}

\begin{figure*}[!t]
\epsscale{1.10}
\plotone{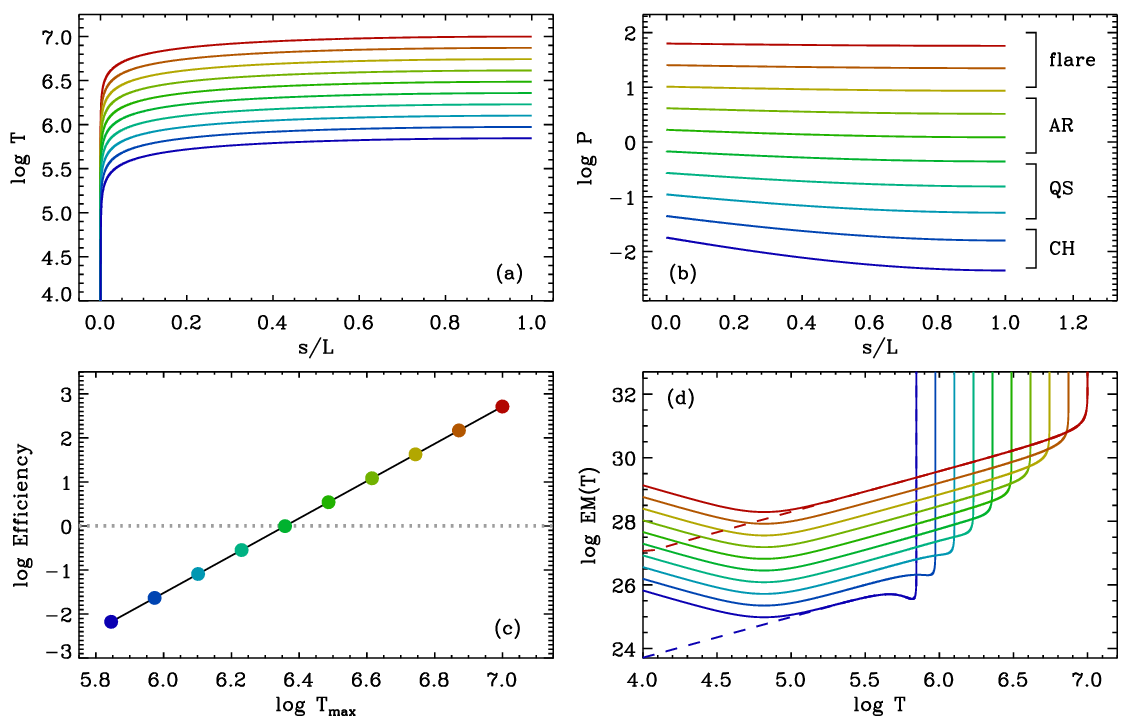}
\caption{Modeled properties for a set of idealized loops,
with color-coding consistent across all panels:
(a) base-10 logarithm of temperature (in K) versus distance
$s$ in units of loop half-length $L$;
(b) base-10 logarithm of pressure (in dyn~cm$^{-2}$) versus $s/L$;
(c) output efficiency factor ${\cal E}$ for each model, with the
critical value ${\cal E}=1$ also shown (dotted line);
(d) base-10 logarithm of EM (in cm$^{-5}$) versus
temperature, with dashed curves showing EM computed with
pure Spitzer-H\"{a}rm conductivity (i.e., no ambipolar diffusion).
Other abbreviations:
CH refers to coronal holes and internetwork regions,
QS refers to ``Quiet Sun'' and chromospheric network,
and AR refers to active regions and plage.
\label{fig08}}
\end{figure*}

First, we chose a representative value of the heating-rate exponents
(e.g., $m=1$ and $n=1$) and fixed the coronal parameters
$L = 100$~Mm and $B_{\odot} = 100$~G.
We also used Equation~(\ref{eq:delfit}) to estimate the $\delta$
parameter, and we selected a grid of ten values of 
$T_{\rm max}$ spaced uniformly between 0.7 and 10 MK.
For each loop in this grid, the peak temperature was thus used as
$T_{\rm norm}$ in order to compute the heating-rate normalization
parameter ${\cal E}_0$ as discussed in Section~\ref{sec:plasma:Ma10}.
This process provided a set of ten example loops that span a range
of likely coronal properties.

Figures~\ref{fig08}(a) and \ref{fig08}(b) show $T(s)$ and $P(s)$
for the ten loops described above.
As shown, the values of $P(s)$ compare favorably to typical
observationally inferred coronal-base pressures for various regions
of the solar atmosphere \citep[e.g.,][]{VAL81,Fo11}.
Figure~\ref{fig08}(c) shows the calculated values of the efficiency
parameter, with an emergent
power-law trend ${\cal E} \propto T_{\rm max}^{4.23}$.
We note that these efficiencies are not automatically
limited to values less than one.
This is not a problem because our standard Poynting flux scaling
for the heating rate (Equation~(\ref{eq:poynt})) was really just
an order-of-magnitude estimate.
Other ways of converting convective kinetic energy into magnetic
energy were described by, e.g.,
\citet{We15}, \citet{Kn18}, and \citet{JK24}.
Also, note that the models exhibiting ${\cal E} \gg 1$ correspond
to active region and flare pressures, and those regions often
have basal magnetic field strengths much larger than the value of
$B_{\odot}$ assumed for this set of models.

Figure~\ref{fig08}(d) shows simulated emission measure
(EM) curves for these loops, both with and without the ambipolar
correction for $\eta$ described in Section~\ref{sec:plasma:cond}.
These were computed by first finding the local differential emission
measure $\mbox{DEM}(T) = n_e^2 (dT/ds)^{-1}$ \citep{Gb76,AS02}.
Then, to best compare with observational data that are typically
plotted over several decades of $T$, we use the integral form
\begin{equation}
  \mbox{EM}(T) \,\, = \,\,
  \int dT \, \frac{n_e^2}{dT/ds} \,\, \approx \,\,
  T \cdot \mbox{DEM}(T)
\end{equation}
\cite[see][]{Wdt86}.
The values at coronal temperatures overlap with observationally
inferred values from the literature, and the fact that $\mbox{EM}(T)$
{\em decreases} as one goes from the upper chromosphere to the low corona 
(i.e., as one goes from 4 to 5 in $\log T$) also agrees with
observational data \citep{Du72,RD81,De82,LC08}.
Traditionally, it has been difficult for models to achieve this
trend with time-steady one-dimensional loop models
\citep{At81,Wdt90}, but the addition of ambipolar diffusion appears
to be one way to remedy that shortcoming.

The behavior of the emission measure in the upper chromosphere and
transition region deserves some additional discussion.
Note that an analytic estimate of its shape can be found by making the
assumption of a spatially constant conductive flux at the coronal
base (i.e., $Q_{\rm cond} \approx 0$; see, e.g.,
\citeauthor{Du72} \citeyear{Du72}).
Considering short spans of loop length $s$, over which the 
magnetic field strength does not vary appreciably, this implies
$\kappa_e \propto (dT/ds)^{-1}$, and thus that
$\mbox{EM} \propto \kappa_e n_e^2 T$.
At the bases of coronal loops, $P \approx$~constant, so $n_e \propto 1/T$,
which leads to a scaling of $\mbox{EM} \propto \kappa_e / T$.
At temperatures where classical Spitzer-H\"{a}rm conduction dominates,
this gives $\mbox{EM} \propto T^{+3/2}$.
However, at lower temperatures where ambipolar diffusion dominates,
this gives $\mbox{EM} \propto T^{-3/2}$.
In this simple model, the transition between these regimes takes place
where $\mbox{EM}$ is minimized, at
$T_{\rm min}/T_{\rm A} = (5^{1/3}) \approx 1.71$.
In the numerical models shown in Figure~\ref{fig08}(d), the minimum
occurs close to this, at about $T_{\rm min}/T_{\rm A} \approx 1.77$.
Thus, our chosen value of $T_{\rm A} =$~40,000~K corresponds to a
minimum at $\log T \approx 4.9$, close to where it is often observed to be.

Lastly, we ought to note that
the models shown in Figure~\ref{fig08} were computed with the
radiative loss parameters (i.e., $\gamma$ and $\chi_0$) for
coronal abundances.
When we recomputed these models for the other (photospheric) set of
abundances described in Section~\ref{sec:plasma:rad}, there were
no major changes to the resulting plasma parameters.
Specifically, the $T(s)$ curves were virtually identical.
The $P(s)$ values were higher for the photospheric abundances,
but only by about 4\% (for the coolest model) to 26\% (for the
hottest model).
The corresponding efficiency values ${\cal E}$ were about 10\% lower.
Thus, we decided to set aside the photospheric values of $\gamma$
and $\chi_0$ and carry out the remainder of the modeling in this
paper using the coronal values.

\subsection{Monte Carlo Ensemble of Heating-Rate Exponents}
\label{sec:params:2nd}

Next, we created a set of models to explore how coronal heating
changes when the $m$ and $n$ exponents in $Q_{\rm heat}$ are allowed
to vary over the ranges proposed by different theories
(see, e.g., Table~\ref{table01}).
Thus, we set up a Monte Carlo ensemble of $10^5$ independent trials,
with each one having unique values of $m$ and $n$ sampled uniformly
between values of 0 and 2.
For each set of these values, we computed ${\cal E}_0$ using the
typical coronal loop of \citet{Sj04} (i.e., $L = 15$~Mm,
$B_{\odot} = 100$~G, $T_{\rm norm} = 3$~MK).
Then for each actual trial loop we used this heating-rate
normalization but chose the loop lengths and basal field strengths
from random distributions meant to reproduce the parameter space
seen in Figure~\ref{fig01}(a).
Specifically, values of $L$ were sampled logarithmically between
10 and 1000~Mm and values of $B_{\odot}$ were sampled logarithmically
between 1 and 1000~G.

\begin{figure}[!t]
\epsscale{1.19}
\plotone{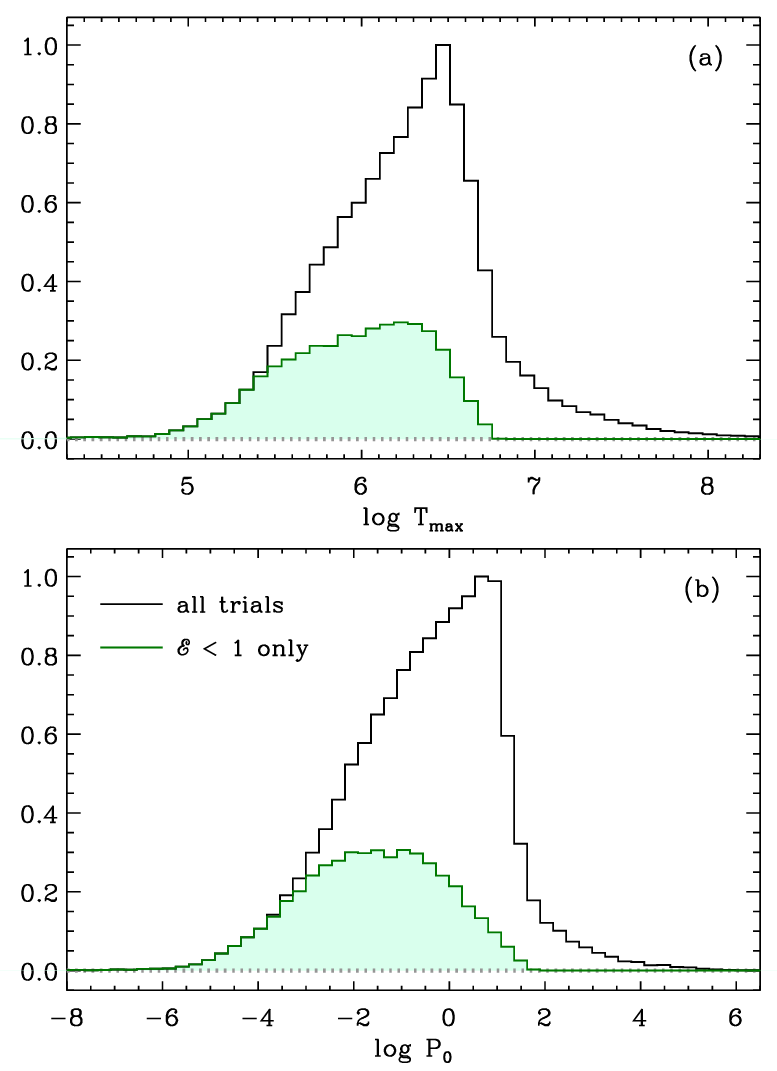}
\caption{Histograms from a Monte Carlo ensemble of coronal loop
models: (a) distribution of values for the
base-10 logarithm of loop-top temperature (in K), and
(b) base-10 logarithm of base pressure (in dyn~cm$^{-2}$), with
the full distribution (solid black curve) compared with the subset
of models with ${\cal E} \leq 1$ (green curve and filled region).
\label{fig09}}
\end{figure}

Figure~\ref{fig09} shows histograms of the resulting statistical
distributions of values for $T_{\rm max}$ and $P_0$.
The ranges are similar to the values seen in Figure~\ref{fig08}
but both distributions have tails that extend both lower and higher.
Out of the full set of Monte Carlo trials, about 35\% of them
exhibited values of the efficiency ratio ${\cal E} \leq 1$, and we show
the distribution of that subset of parameters as a separate histogram.
Median values for $\log T_{\rm max}$ were 6.28 for the full set
and 5.99 for the subset with ${\cal E} \leq 1$.
Median values for $\log P_0$ were --0.36 for the full set
and --1.58 for the subset with ${\cal E} \leq 1$.

To illustrate the effects of varying $m$ and $n$ while
keeping the normalization procedure (for ${\cal E}_0$) the same
for all models, we examined the median values of $T_{\rm max}$
and $P_0$ for sixteen subsets of the Monte Carlo ensemble.
Table~\ref{table02} displays these values in 4$\times$4 grids.
The subsets are defined by selecting independent ranges of each of
the $m$ and $n$ exponents (i.e., 0--0.5, 0.5--1, 1--1.5, or 1.5--2),
and there are roughly 6,250 trial models in each subset.
Note that, despite the consistent normalization for ${\cal E}_0$,
the models with the lowest values of $m$ and the highest
values of $n$ tend to produce distributions of coronal loops
with relatively ``quiet'' (i.e., lower) temperatures and pressures.
On the other hand, models with the highest values of $m$ and the
lowest values of $n$ tend to have distributions weighted more
toward ``active'' plasma conditions.

The trends seen in Table~\ref{table02} are not immediately apparent from
the heating-rate scaling laws given in Section~\ref{sec:heating:mine}.
For example, Equation~(\ref{eq:Qheatvsmn}) indicates that
smaller $m$ exponents are correlated with heating rates that vary
strongly with the magnetic field strength, whereas larger values of
$m$ indicate a weaker dependence on $B$.
Given that high solar activity tends to go along with the largest
dynamic range (and largest maximum values) in coronal emission,
one may have assumed that models with low values of $m$ would
produce loops with the largest values of $T_{\rm max}$ and $P_0$.
However, note that the $m$ exponent also constrains the spatial
dependence of $Q_{\rm heat}$.
For reasonable values of the $\delta$ exponent, smaller values of $m$ 
correspond to smaller ``heating rate scale heights,''
and these often lead to lower values of $T_{\rm max}$
(see, e.g., Equation~(37) of \citeauthor{AS02} \citeyear{AS02}).

\begin{table*}
\caption{Median Coronal Parameters from Monte Carlo Ensemble
\label{table02}}
\hspace*{0.97in}
\begin{tabular}{lcccc}
\hline
\hline
$\log T_{\rm max}$ & & & \\
          & $0 \leq m < 0.5$ & $0.5 \leq m < 1$
          & $1 \leq m < 1.5$ & $1.5 \leq m < 2$ \\
$0 \leq n < 0.5$ & 6.150     & 6.315     & 6.540     & 6.964 \\
$0.5 \leq n < 1$ & 6.050     & 6.186     & 6.397     & 6.757 \\
$1 \leq n < 1.5$ & 5.907     & 6.042     & 6.244     & 6.547 \\
$1.5 \leq n < 2$ & 5.815     & 5.912     & 6.069     & 6.393 \\
\hline
$\log P_{0}$ & & & \\
          & $0 \leq m < 0.5$ & $0.5 \leq m < 1$
          & $1 \leq m < 1.5$ & $1.5 \leq m < 2$ \\
$0 \leq n < 0.5$ & --0.913   & --0.379   &   0.409   &   1.611 \\
$0.5 \leq n <1 $ & --1.208   & --0.765   & --0.092   &   1.042 \\
$1 \leq n < 1.5$ & --1.582   & --1.147   & --0.570   &   0.575 \\
$1.5 \leq n < 2$ & --1.843   & --1.514   & --1.046   & --0.041 \\
\hline
\end{tabular}
\end{table*}

\subsection{Spatial Dependence Along the Line of Sight}
\label{sec:params:3rd}

\begin{figure}[!t]
\epsscale{1.19}
\plotone{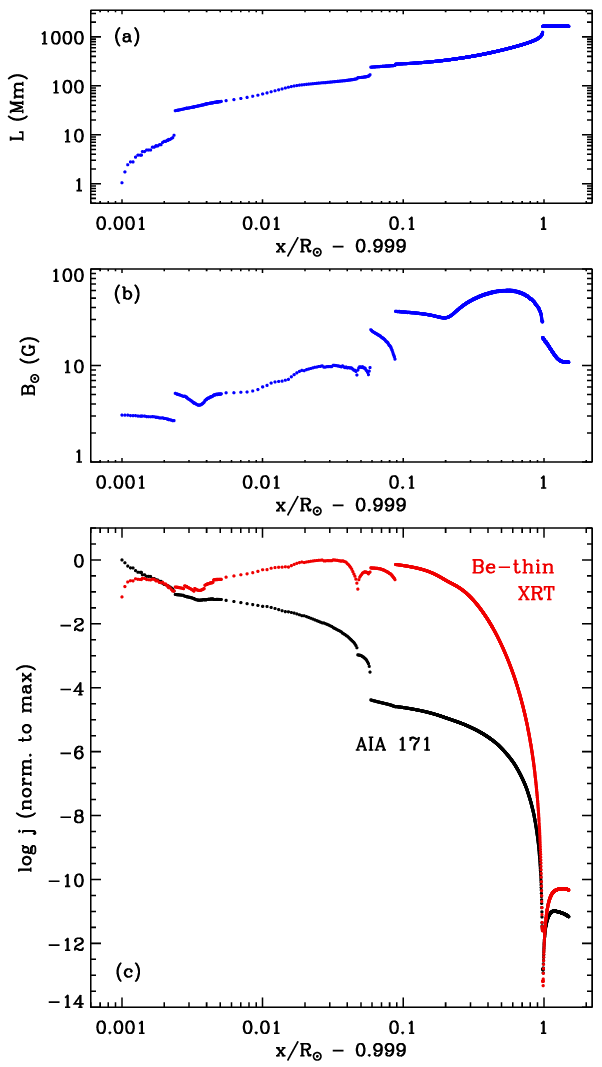}
\caption{Example LOS dependence of (a) loop length $L$,
(b) basal field strength $B_{\odot}$, and
(c) optically thin emissivities for the AIA 17.1~nm filter
(black points) and the XRT Be-thin filter (red points).
To enable visual inspection of the smallest features near the
photosphere, the distance along the LOS is shown as the difference
between Cartesian coordinate $x$ and a fixed value of 0.999~$R_{\odot}$.
Emissivities are shown as the base-10 logarithm of the ratio of
$j_i(x)$ to its maximum value along this particular LOS ray.
\label{fig10}}
\end{figure}

Finally, we chose to examine the LOS dependence of the plasma
properties for one example instance of a three-dimensional corona.
When simulating EUV and X-ray emission for a given observing ray,
the spatial dependence along each loop's magnetic field (i.e.,
along $s$) is not as important as the larger-scale variation of
plasma parameters parallel to the $x$-axis.
Figures~\ref{fig10}(a) and \ref{fig10}(b) show how two key
input parameters ($L$ and $B_{\odot}$) vary when sampled along a ray
that intersects the center of the Sun and points directly toward an
observer along the heliographic equator.
Each point along the LOS intersects a different field line, so
we expect to see point-to-point discontinuities along this ray
when the global connectivity of the magnetic field changes.
Note that the values of $L > 1000$~Mm correspond to points
intersecting open field lines.
Figure~\ref{fig10}(c) shows two examples of filter-specific
emissivities $j_i(x)$ along this ray, which are defined in
Equation~(\ref{eq:intensity}).

For this example calculation, we chose a disk-center LOS for the
solar maximum date 2013 December 20, 20:00~UT.
In this case, the $x$ coordinate extends from the coronal base to the
source surface---i.e., from $x = 1$ to $2.5 \, R_{\odot}$---and
both $y$ and $z$ coordinates were held fixed at zero.
We made the choice to not extend the ray beyond the source surface,
but in the future we will need to reexamine this for image synthesis
with fields of view larger than those of AIA and XRT.
To accurately simulate the corona, the spatial grid must be fine enough
not to miss any important topological magnetic-field variations
along the ray.
Thus, for this example calculation we used a finely spaced (but uneven)
grid of roughly 1400 points in $x$ that was constructed iteratively
to ensure no magnetic structures were missed.
The LOS grid spacing varied with distance; at the base, we used
$\Delta x = 5 \times 10^{-5} \, R_{\odot}$, and at the largest height it
expanded to no larger than $\Delta x = 1.5 \times 10^{-3} \, R_{\odot}$.

For the coronal heating model, we assumed $m=1$, $n=1$, the
radiative $\gamma$ and $\chi_0$ parameters for coronal abundances,
and the same normalization for ${\cal E}_0$ that was discussed in
Section~\ref{sec:params:2nd}.
For the temperature response functions of the filters, we assumed
$t_n = 0$ and used the time-degradation calibrations appropriate for
2013 December 20.
Note that ``cooler'' filters like AIA 17.1~nm exhibit emission that
is strongest near the coronal base, whereas hotter filters (like all
of the XRT bands) generate emission over a more extended range of
heights closer to the loop-tops.
In this case, the neglect of regions above the source surface is
justified because the open field lines sampled at the top of the
observing ray (i.e., for $x > 1.97 \, R_{\odot}$) give rise to
negligibly weak EUV and X-ray emission compared to the loops at
lower heights.

The high-resolution grid used for Figure~\ref{fig10} helped us
optimize the parameters for the coarser grid used in the more
computationally intensive sets of models discussed below.
At the base, we decided to retain the high resolution from this
example grid, but for larger heights we found that larger LOS steps
were acceptable.
Thus, for each LOS ray to be sampled in Section~\ref{sec:results},
we specified a grid along the $x$-axis with 300 points, and we
used radially dependent spacing that expands linearly with distance
above the photosphere.
For example, a disk-center ray (with a total LOS distance of
1.5~$R_{\odot}$) has a grid with
$\Delta x = 1.68 \times 10^{-5} \, R_{\odot}$ at the base and
$\Delta x = 1.01 \times 10^{-2} \, R_{\odot}$ at the source surface.
Rays near the limb have spacings that are larger at all points by
about 53\% due to their spanning a larger LOS distance of
$(2.5^2 - 1)^{1/2} \approx 2.29$~$R_\odot$.

\section{Results: Coronal Heating at Solar Minimum and Maximum}
\label{sec:results}

The primary results of this paper involve an in-depth comparison
between observed and simulated EUV and X-ray intensities for the
two chosen time periods that are meant to be representative of
activity levels appropriate for solar minimum (2018) and
solar maximum (2013).

\subsection{Detailed Comparison with AIA and XRT Data}
\label{sec:results:obs}

In Section~\ref{sec:emission:obs} we described the grids of observing
rays used for the two epochs, and in Section~\ref{sec:emission:sim}
we discussed the specific AIA and XRT filters included in the analysis.
The total numbers of data points for each of the four main cases (solar
minimum vs.\  maximum, and using all data vs.\  excluding active
regions) are provided in the four columns of Table~\ref{table03}.
When evaluating the measured and simulated intensities, we realize
that absolute intensity calibrations---both for the data and for
the $K_i(T)$ response functions---often have substantially higher
uncertainty levels than do ratios.
Thus, we refrain from making absolute comparisons using DN and instead
choose to compare intensity ratios of the form
${\cal R}_{ij} = I_i / I_j$.
Some of these ratios compare one observing ray to another (but for the
same filter), and others compare one filter to another (for the same
instrument, but either for different rays or for the same ray).

\begin{table*}
\caption{Optimum Model Parameter Values
\label{table03}}
\hspace*{0.80in}
\begin{tabular}{lcccc}
\hline
\hline
Quantity & 2018 (all) & 2018 (no AR) & 2013 (all) & 2013 (no AR) \\
\hline
Num.\  LOS rays      & 22  & 21  & 25  & 19  \\
Num.\  AIA filters   & 6   & 6   & 6   & 6   \\
Num.\  XRT filters   & 4   & 4   & 6   & 6   \\
Num.\  data points   & 218 & 208 & 297 & 225 \\
\hline
Fixed $t_n = 0$ & & & \\
\hspace*{0.15in}$\min(\chi^2)$  & 0.1730 & 0.1232 & 0.2585 & 0.1630 \\
\hspace*{0.15in}best $m$  &       1.607  & 1.731  & 1.255  & 1.563 \\
\hspace*{0.15in}best $n$  &       1.144  & 1.310  & 0.6428 & 0.9625 \\
\hspace*{0.15in}best $T_{\rm norm}$ (MK) & 
                                  1.487  & 1.197  & 2.922  & 1.829 \\
\hline
Optimized $t_n$ & & & \\
\hspace*{0.15in}$\min(\chi^2)$  & 0.1513 & 0.1072 & 0.2585 & 0.1549 \\
\hspace*{0.15in}best $m$  &       1.630  & 1.789  & 1.255  & 1.639 \\
\hspace*{0.15in}best $n$  &       1.221  & 1.440  & 0.6428 & 1.054 \\
\hspace*{0.15in}best $T_{\rm norm}$ (MK) &
                                  0.8259 & 0.6292 & 2.922  & 0.8480 \\
\hspace*{0.15in}best $t_n$ (s) &  2080   & 2080   & 0      & 1880 \\
\hspace*{0.15in}best $a=2-m$ &    0.370  & 0.211  & 0.745  & 0.361 \\
\hspace*{0.15in}best $b=1+n-m$ &  0.591  & 0.651  & 0.388  & 0.415 \\
\hline
\end{tabular}
\end{table*}

Thus, for a given year (2013 or 2018) and a given instrument (AIA or XRT)
there is one ray and one filter chosen to be a common denominator $I_j$.
We then loop over all other combinations of rays and filters for $I_i$
when we determine a normalized goodness-of-fit parameter,
\begin{equation}
  \chi^2 \, = \, \frac{1}{N} \sum_i
  \left[ \log_{10} ( {\cal R}_{ij,{\rm mod}} ) -
         \log_{10} ( {\cal R}_{ij,{\rm obs}} ) \right]^2 \,\,\, ,
\end{equation}
where $N$ is the number of data points for a given case (see
Table~\ref{table03}) and subscripts ``mod'' and ``obs'' refer to
modeled and observed values, respectively.
Because we use one AIA measurement for the denominator $I_j$ when
comparing with other AIA data points, and one XRT measurement for the
denominator when comparing with other XRT data points, we usually find
that $N$ is given by two less than the product of the number of LOS rays
and available filters.
However, for the solar maximum epoch (2013) there was one unusable data
point from the XRT Al-med filter, so $N$ is given by three less than the
product of the number of LOS rays and available filters.
Taking both years into account, we have 515 individual intensity
ratios to compare between the models and observations.

For our modeling procedure, we independently varied four
free parameters that describe the coronal-heating physics:
the $m$ and $n$ exponents, the $T_{\rm norm}$ normalization constant
for ${\cal E}_0$, and the DEM broadening timescale $t_n$.
Generally, the first three of those parameters were treated
together because the simulation of three-dimensional plasma properties
(i.e., densities and temperatures at each point along each ray)
depended on all of them.
The fourth parameter ($t_n$) was used only at the later stage of
integrating emissivities along each LOS ray, so it did not need to
be varied when computing the spatially-dependent plasma properties.
All of the models discussed in this section used the $\gamma$ and
$\chi_0$ values for coronal abundances.

\begin{figure}[!t]
\epsscale{1.19}
\plotone{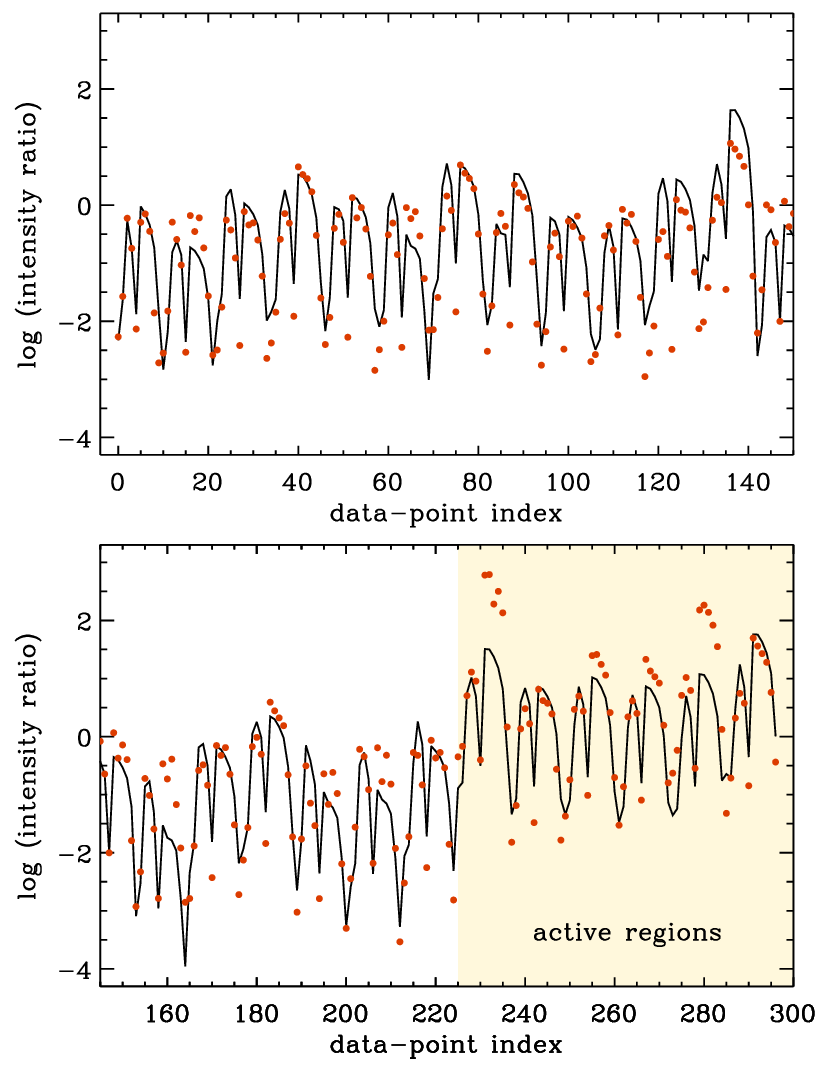}
\caption{Comparison between the full set of ratios ${\cal R}_{ij}$
for a model of the 2013 corona (black curves; see text for details)
and the observed 2013 data from AIA and XRT (red points).
The ``data-point index'' first cycles through the AIA and XRT
ratios for 19 observing rays in a hexagonal pattern (white background),
then for the remaining 6 rays that align with active regions
(gold background).
The two panels split up the full set of ratios into sequential groups.
\label{fig11}}
\end{figure}

Figure~\ref{fig11} shows an example comparison between the 297 observed
and modeled data points for the 2013 (solar maximum) case.
As shown in Table~\ref{table03} and discussed above, the observations
for this case have either 11 or 12 data points for each ray.
Figure~\ref{fig11} cycles through each set, going from left to right,
first for the 19 rays corresponding to the hexagonal pattern shown in
Figure~\ref{fig05}, then for the 6 additional rays chosen to align
with active regions.
Because the intensities used for the denominators in ${\cal R}_{ij}$
are always chosen from the hexagonal grid and not the active regions,
the ratios for the latter are generally higher than those for the former.
This can be seen clearly in the region with the gold background.

The model parameters for the comparison shown in Figure~\ref{fig11}
were given by $t_n = 0$, $m = 1.2$, $n = 0.5$, and
$T_{\rm norm} = 2.86$~MK.
These were the parameters that gave rise to the lowest $\chi^2$
parameter for a coarse three-dimensional grid of
$m$, $n$, and $T_{\rm norm}$ values, and $t_n$ was fixed at zero.
This grid contained 55 values of $m$ (with $-0.2 \leq m \leq 2.5$),
101 values of $n$ (with $-2.5 \leq n \leq 2.5$), and
20 values of $T_{\rm norm}$ (with $0.8 \leq T_{\rm norm} \leq 6$~MK)
for a total of 111,100 independent models.
For the best-fitting set of parameters given above,
the resulting minimum value of $\chi^2$ was 0.259 for the comparison
involving all 297 data points.
For just the subset of 225 non-active-region rays, the agreement was
better, with $\chi^2 = 0.163$.

\begin{figure}[!t]
\epsscale{1.17}
\plotone{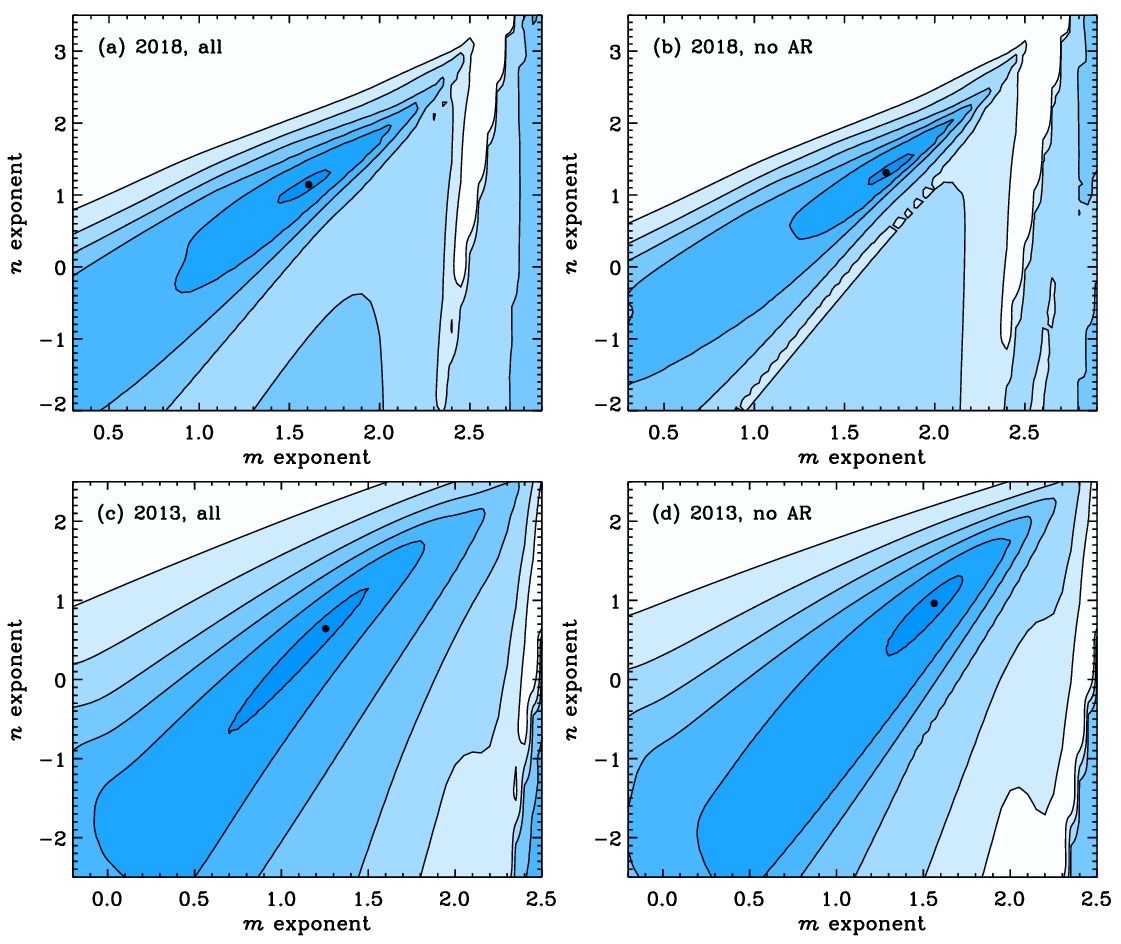}
\caption{Contours of $\chi^2$ for
(a) the solar minimum comparison with all data,
(b) the solar minimum comparison without active regions,
(c) the solar maximum comparison with all data,
(d) the solar maximum comparison without active regions.
Absolute minimum values (black circles) are
surrounded by contours filled with successively brighter colors,
each one spaced by multiplicative factors of two.
Contour values of $\chi^2$ start at 1.15 times the absolute
minimum and end at 36.8 times the minimum (i.e., $1.15 \times 32$).
\label{fig12}}
\end{figure}

Next, Figure~\ref{fig12} shows contours of $\chi^2$
as a function of $m$ and $n$ for the four primary cases corresponding
to the columns of Table~\ref{table03}.
These plots are the result of the same three-dimensional grid of models
discussed above, but for each point in the two-dimensional space shown in
the figure, we chose the value of $T_{\rm norm}$ that had the lowest $\chi^2$.
As before, we kept the $t_n$ parameter fixed at zero for this case.
In all four panels of Figure~\ref{fig12}, it looks like $\chi^2$ is
beginning to decrease to the right, as $m$ increases past values of 2.5.
However, this region of parameter space is not generally favored
because $m > 2$ would be consistent with the heating rate getting
larger for a {\em decreasing} magnetic field.

\begin{figure}[!t]
\epsscale{1.19}
\plotone{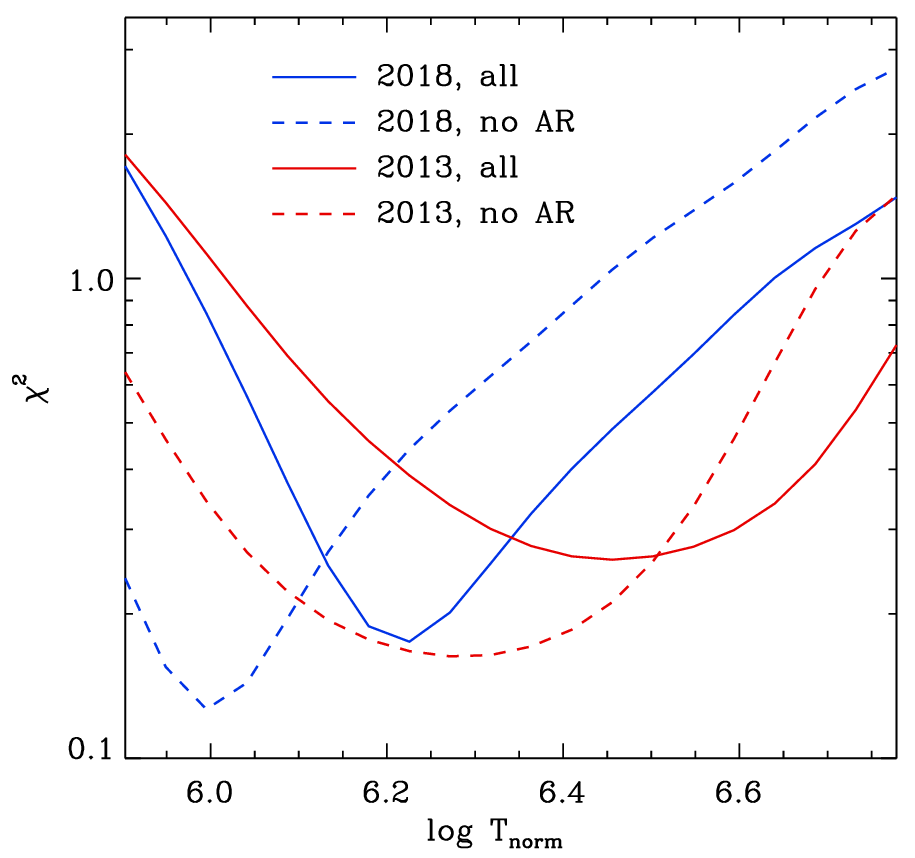}
\caption{Dependence of $\chi^2$ on $T_{\rm norm}$ for the 
absolute-minimum points shown in the four panels of Figure~\ref{fig12}.
Comparisons with solar-minimum data (blue curves) and solar-maximum
data (red curves) are shown for the cases of including all modeled
rays (solid curves) and excluding active regions (dashed curves).
\label{fig13}}
\end{figure}

Because Figure~\ref{fig12} does not illustrate how $\chi^2$ varies
as a function of $T_{\rm norm}$, we show some examples of this behavior
in Figure~\ref{fig13}.
Specifically, this provides the $T_{\rm norm}$ dependence of $\chi^2$
for each of the four choices of $m$ and $n$ corresponding to the
absolute minima shown in Figure~\ref{fig12}.
All four curves show distinct minima in the range of parameter space
chosen for exploration for this heating-rate normalization parameter.
There is also a general trend for the optimal values of $T_{\rm norm}$ 
to increase as a function of overall solar activity.
In other words, the solar-minimum data without active regions
has the lowest value of $T_{\rm norm}$, and the
solar-maximum data with active regions has the highest value.
Note that the least active of these four cases has the highest
values of both $m$ and $n$, and the most active has the lowest values
of $m$ and $n$.
We discuss the implications of correlations like this in
Section~\ref{sec:conc}.

After computing the three-dimensional grid of $\chi^2$ values
(as a function of $m$, $n$, and $T_{\rm norm}$, for a fixed value
of $t_n = 0$), we subsequently ran randomized Monte Carlo searches
that were focused on the global minima regions shown in
Figures~\ref{fig12} and \ref{fig13},
By creating several hundred new combinations of the three input
parameters, these models allowed us to zoom in on the ``sub-grid scale''
global minimum region of $\chi^2$ space and thus determine the optimal
values of $m$, $n$, and $T_{\rm norm}$ to a higher degree of accuracy.
The middle section of Table~\ref{table03} displays these more precise
values, still for the $t_n = 0$ case.

\begin{figure}[!t]
\epsscale{1.19}
\plotone{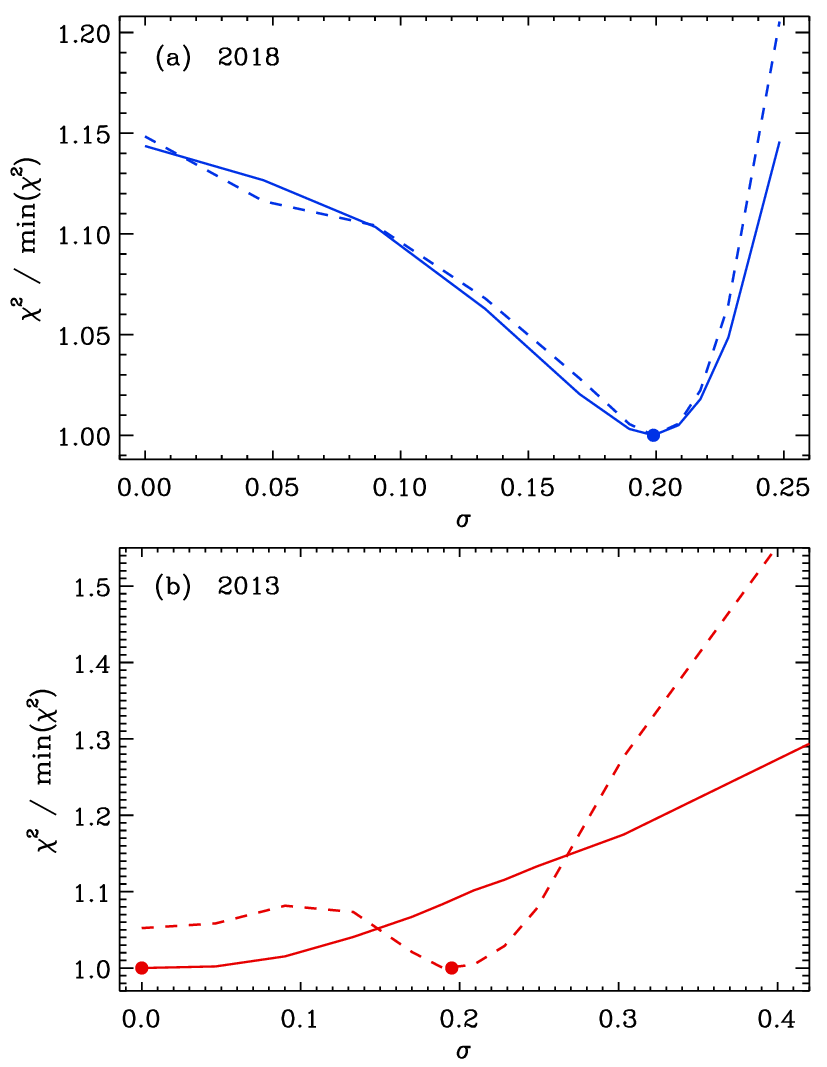}
\caption{Dependence of $\chi^2$ (normalized by its minimum value)
on $t_n$ for the four observation-comparison cases discussed above.
The horizontal axis is shown as $\sigma$, which is related to $t_n$ via
Equation~(\ref{eq:sigmadef}).
The curves have the same colors and styles as in Figure~\ref{fig13}.
\label{fig14}}
\end{figure}

Lastly, we needed to vary the $t_n$ parameter associated with the
width of the multithermal DEM from nanoflare activity.
The bottom section of Table~\ref{table03} provides the optimal values
of $m$, $n$, and $T_{\rm norm}$ found when $t_n$ was allowed to
be varied freely as well.
For three out of the four cases, a nonzero value of $t_n$ gave better
agreement with the data than did the assumption of $t_n = 0$.
Figure~\ref{fig14} shows how $\chi^2$ varies as a function of $t_n$
(or, equivalently, $\sigma$) for each of the four cases discussed above,
with the optimal values of $m$, $n$, and $T_{\rm norm}$ held fixed
for each case.
For these results, Table~\ref{table03} also gives the optimal values
of the $a$ and $b$ parameters defined in Equation~(\ref{eq:abdef}).
In all cases, both parameters tend to fall between 0 and 1, which
agrees with many past observational inferences
(see, e.g., Table~4 of \citeauthor{Is24} \citeyear{Is24}).

\subsection{Discussion of Best-Fitting Heating Rate Exponents}
\label{sec:results:disc}

\begin{figure}[!t]
\epsscale{1.19}
\plotone{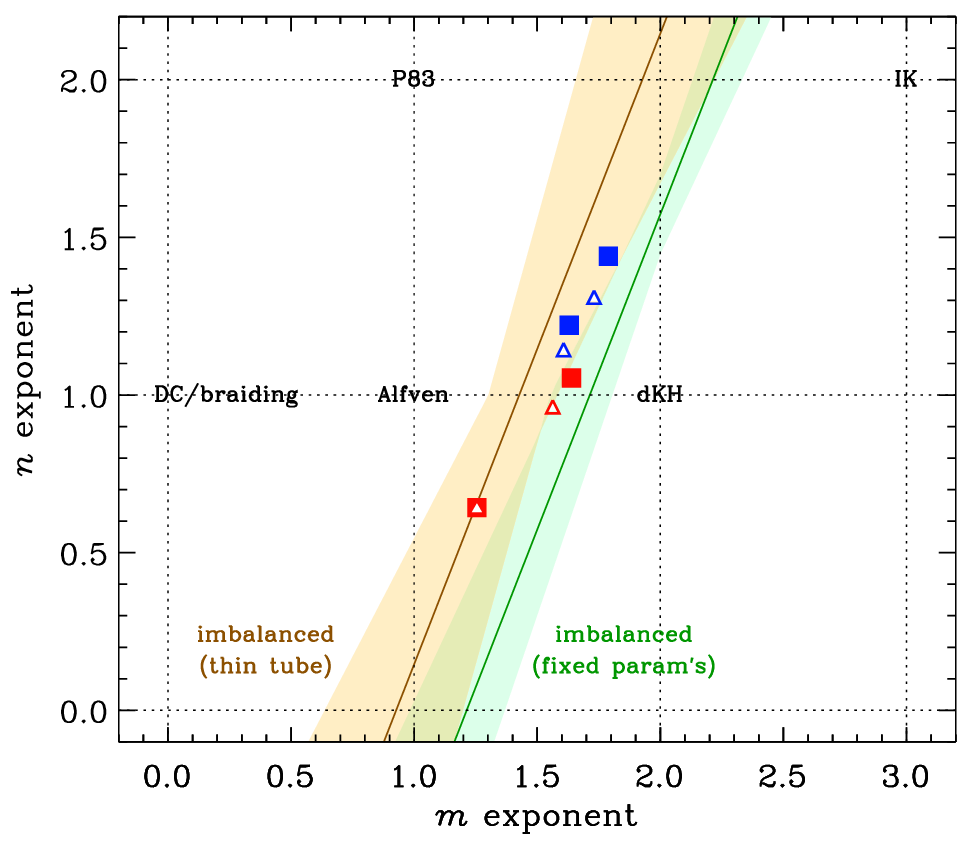}
\caption{Parameter summary of heating-rate exponents that
best match the AIA and XRT data.
For the points, colors match those shown in earlier plots
(i.e., blue for 2018, red for 2013) and symbols indicate
whether the best value of $t_n$ was used (filled squares) 
or $t_n = 0$ was used (open triangles).
Other locations in the parameter space are labeled by, e.g.,
P83 (for \citeauthor{P83} \citeyear{P83}),
dKH \citep{dKH38}, and IK (Iroshnikov \& Kraichnan).
Power-law trends for imbalanced MHD turbulence are shown for
fixed values of the basal parameters (green line/region)
and spatial dependences following the thin-tube approximation
(light brown line/region).
\label{fig15}}
\end{figure}

Figure~\ref{fig15} illustrates the parameter space defined by the $m$
and $n$ heating exponents, with the red and blue points indicating
$\chi^2$-minimum values from the eight cases listed in
Table~\ref{table03}.
Using only the four cases that allowed $t_n$ to vary freely (i.e., the
square symbols only), we fit a linear function to the mutual dependence
between the two exponents and obtained $n \approx 1.436m - 1.176$.
Note that more active data-sets seem to be explained best by lower values
of both $m$ and $n$, reminiscent of the DC/nanoflare
mechanisms that reside in the lower-left part of the diagram.
On the other hand, more quiet data-sets correspond to higher values of
both exponents, more consistent with AC (wave/turbulence) mechanisms
that appear in the upper-right part of the diagram.
Our values of $m$ and $n$ fall in between the cases
of a heating rate that scales with the Alfv\'{e}n-wave energy flux
($m=1$, $n=1$) and one that agrees with Kolmogorov turbulence
($m=2$, $n=1$; more specifically, the constant-energy-flux scaling
developed by \citeauthor{dKH38} \citeyear{dKH38}).

Because the empirically determined values of $m$ and $n$ are
somewhat close to the prediction for standard hydrodynamic turbulence,
it is worthwhile to investigate how those predicted values would be
modified for {\em imbalanced} turbulence in the strong-field
(i.e., low-plasma-$\beta$) environment of coronal loops.
Imbalanced turbulence is most often invoked for open field lines
connected to the solar wind, since in that case there is always a
clear dominance of outward-propagating fluctuations over
inward-propagating fluctuations.
However, even though closed loops have sources of waves at both ends,
the existence of localized dissipation and large-scale asymmetries
means that the fluxes of waves in both directions will not tend to
remain in balance at all points \citep[see, e.g.,][]{vB11,Shi24}.

For turbulence made up of transverse Alfv\'{e}nic fluctuations
(i.e., perpendicular oscillations in velocity $v_{\perp}$
correlated with perpendicular oscillations in the magnetic field
$B_{\perp}$), the expected heating rate for imbalanced turbulence can
be written in terms of Elsasser variables,
\begin{equation}
  z_{\pm} \, = \, v_{\perp} \mp \frac{B_{\perp}}{\sqrt{4\pi\rho}}
\end{equation}
\citep{E50}, where the two signs represent wave packets propagating
along the field in opposite directions.
Generally we use $z_{+}$ for the amplitude of the dominant mode
with larger energy density at a given location, and $z_{-}$ refers to
the less energetic mode.
If we define a dimensionless Elsasser imbalance ratio
${\cal I} = |z_{-}/z_{+}|$, we can use a standard phenomenological
expression for the quasi-steady heating rate,
\begin{equation}
  Q_{\rm heat} \, = \,
  \frac{\rho (z_{+} z_{-}^2 + z_{+}^2 z_{-})}{\lambda_{\perp}}
  \, \approx \, \frac{\rho v_{\perp}^3}{\lambda_{\perp}} \, f({\cal I})
  \label{eq:QheatfI}
\end{equation}
\citep[e.g.,][]{Hs95,Mt99}.
In this expression, if we make standard assumptions about the
Alfv\'{e}nicity of the wave packets and the long-term ensemble
averaging of $z_{\pm}$, we can write
\begin{equation}
  f({\cal I}) \, = \, \frac{{\cal I} ({\cal I} + 1)}{2} \,\, ,
  \label{eq:fIdef}
\end{equation}
which simplifies the problem by approximating the Kolmogorov limit
as equivalent to perfectly balanced turbulence (${\cal I}=1$).
In addition, the scaling relations defined in
Section~\ref{sec:heating:mine} can be used to rewrite
Equation~(\ref{eq:QheatfI}) as
\begin{equation}
  f({\cal I}) \, \propto \, 
  \left( \frac{\tau_{\rm A}}{\tau_{\rm ph}} \right)^{m-2}
  \left( \frac{\lambda_{\perp}}{L} \right)^{n-1}
  \,\,\, .
  \label{eq:fIscaling}
\end{equation}
Examining how the left-hand and right-hand sides of
Equation~(\ref{eq:fIscaling}) depend on loop-length $L$ will lead
to specific interrelationships between $m$ and $n$ for imbalanced
turbulence.

\begin{figure}[!t]
\epsscale{1.17}
\plotone{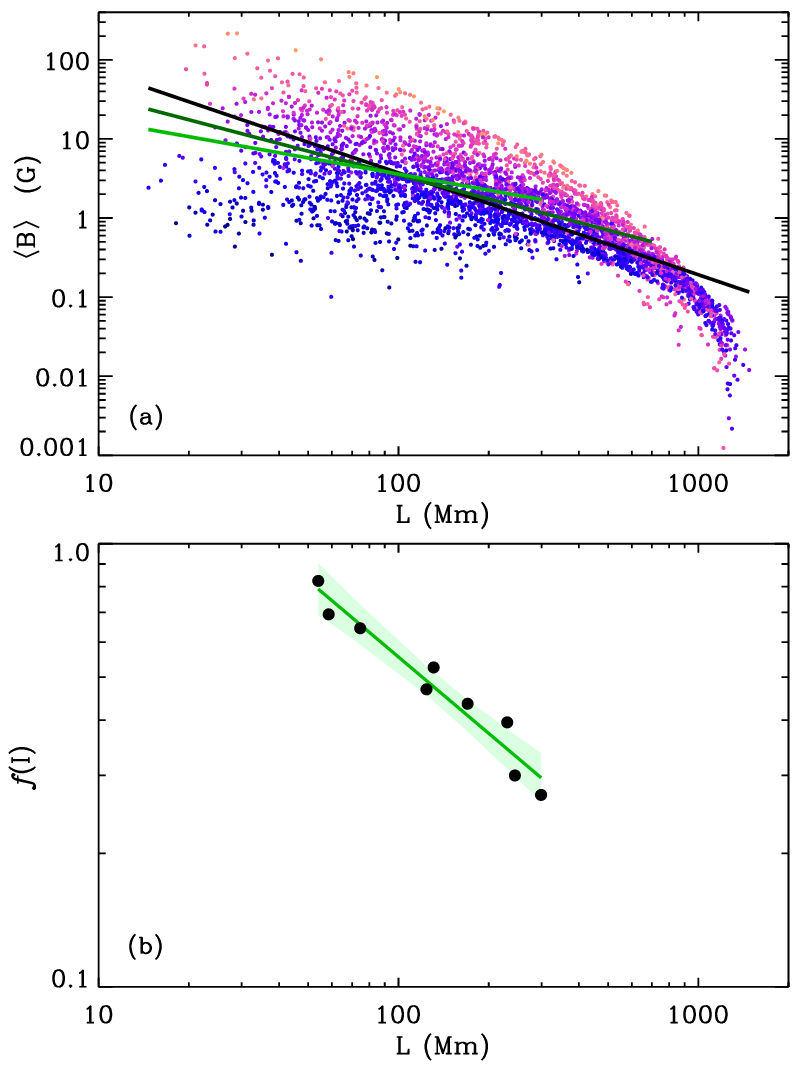}
\caption{(a) Numerically computed values of $\langle B \rangle$
from Equation~(\ref{eq:Bmean}) with point-colors corresponding
to the data shown in Figure~\ref{fig01}.
Power-law fits are shown for three cases:
including all $L$ values (black line, $k = -1.29$),
loops with $L \leq 700$~Mm (dark green line, $k = -0.99$), and
loops with $L \leq 300$~Mm (medium green line, $k = -0.67$).
(b) Turbulence imbalance factors $f({\cal I})$ extracted from
simulations \citep{Shi24} (black points), a power-law fit
(medium green line, $r = -0.573$), and 95\% confidence interval
(light green region).
\label{fig16}}
\end{figure}

For the right-hand side of Equation~(\ref{eq:fIscaling}), we note that
the local magnetic field strength $B$ appears in $\tau_{\rm A}$.
Although we saw no strong dependence of the basal field strength
$B_{\odot}$ on $L$ (see Figure~\ref{fig01}), it must be the case that
the field-line-averaged value becomes weaker for longer loops.
Figure~\ref{fig16}(a) shows the result of calculating
\begin{equation}
  \langle B \rangle \, = \, \frac{1}{L} \int_0^L ds \,\, B(s)
  \label{eq:Bmean}
\end{equation}
using the 5,000 field lines shown in Figure~\ref{fig01}, together
with Equation~(\ref{eq:deltadef}) and the incomplete beta function
solution for $T(s)$ discussed in Section~\ref{sec:plasma:spatial}.
Based on the data shown in Figure~\ref{fig16}(a), we assumed a
power-law fit (i.e., a linear fit in log-log space)
for $\langle B \rangle \propto L^{k}$.
Performing such a fit with various subsets of the data gives a range
of values for $k$ between --1.29 and --0.67.
An intermediate value of $k \approx -1$ essentially agrees with the
observational results of \citet{WW06}, who made use of the scaling
relation $\langle B \rangle \propto B_{\odot} / L$.

For the left-hand side of Equation~(\ref{eq:fIscaling}), we used
numerical simulations of MHD turbulence to determine how the
imbalance factor $f({\cal I})$ depends on loop length $L$.
Examples of these kinds of simulations were presented in Figure~10
of \citet{Shi24} for nine loops with corresponding values of $L$
between 50 and 300 Mm.
These simulations showed that longer loops tend to exhibit more
imbalanced turbulence than shorter loops, and
additional examples from simulations \citep{Do16} and
observations \citep{Ta19} can be found.

We extracted values for ${\cal I}$ from three locations
along each of the loops from \citet{Shi24} (the two coronal bases and
the midpoint), computed averages for each loop, and used
Equation~(\ref{eq:fIdef}) to determine $f({\cal I})$ for each case.
Figure~\ref{fig16}(b) shows that a power-law fit to these data points
gives $f({\cal I}) \propto L^{r}$, with $r = -0.573$.
Because this fit is based on only nine values (each having been sampled
very coarsely from the published plots), we determined a confidence
interval on the slope using the ordinary least squares (OLS)
routine from the Python {\tt statsmodels} package \citep{SP10}.
Assuming normally distributed uncertainties, the 95\%
confidence interval on the power-law exponent is $-0.698 < r < -0.447$.

With the above analysis of the left-hand and right-hand sides of
Equation~(\ref{eq:fIscaling}) in place, we can compare them by making
two alternate sets of assumptions:
\begin{enumerate}
\item
First, if we follow the discussion in Section~\ref{sec:heating:mine}
and consider the quantities $\lambda_{\perp}$, $v_{\perp}$, and
$\rho$ to be fixed constants, this means that
\begin{equation}
  f({\cal I}) \, \propto \, B^{2-m} L^{m-n-1} \,\,\, .
  \label{eq:fIBL}
\end{equation}
Note that this is a somewhat cavalier simplification, since we know
that $\rho$ varies with distance $s$ for different types of loops
and $v_{\perp}$ varies between regions of weak and strong background
field strength.
However, outside of the largest sunspots, these changes in $v_{\perp}$
can be relatively small and the dependence of $Q_{\rm heat}$ on $\rho$
is often weaker than the dependences on the other primary quantities.
Using the power-law fits for $\langle B \rangle$ and
$f({\cal I})$, Equation~(\ref{eq:fIBL}) gives
\begin{equation}
  r \, = \, (2-m)k + m - n - 1
  \label{eq:krmn1}
\end{equation}
which provides a theoretical prediction for a linear locus of
points in the $(m,n)$ plane that correspond to imbalanced turbulence.
Figure~\ref{fig15} shows the outline defined by the collection of
linear curves that result from inserting the above ranges for
$k$ and $r$ into Equation~(\ref{eq:krmn1}), then solving for $n(m)$.
For the most likely values ($k=-1$ and $r=-0.573$), the central
linear trend is $n = 2m - 2.427$.
\item
Alternately, we can interpret $\lambda_{\perp}$, $v_{\perp}$, and
$\rho$ as spatially dependent quantities.
In thin flux-tube theory (generally most applicable in the upper
photosphere and chromosphere; see \citeauthor{Sp81} \citeyear{Sp81}),
one can assume a balance between gas pressure in an isothermal
atmosphere and magnetic pressure to see that $\rho \propto B^2$.
For energy-conserving Alfv\'{e}nic fluctuations, we can also assume
$v_{\perp} \propto \rho^{-1/4} \propto B^{-1/2}$.
Also, if the transverse scale-length for horizontal motions remains
proportional to the width of the flux-tube, we can assume
$\lambda_{\perp} \propto B^{-1/2}$.
\citet{Cr09} showed how these scalings combine so that the classical
turbulent heating rate \citep{dKH38} varies linearly with $B$.
With these scalings,
\begin{equation}
  r \, = \, m - n - 1 + \frac{k}{2}(1-n)
  \label{eq:krmn2}
\end{equation}
and Figure~\ref{fig15} shows an outline for the resulting
range of $n(m)$ curves.
For the central values ($k = -1$ and $r = -0.573$), the relationship
is $n = 2m - 1.854$, and it has the same
slope ($n \propto 2m$) as the above model for the fixed parameters.
\end{enumerate}
As we see in Figure~\ref{fig15}, the empirically derived values of
$m$ and $n$ from Table~\ref{table03} fall very close to the
overlap region between these two interpretations of the imbalanced
turbulence model.
The slope of the fit to the best set of empirically derived
$m$ and $n$ parameters (i.e., $n \propto 1.436 m$) is slightly
shallower than the central trends for the two model interpretations,
but the points fall inside the uncertainty bounds of the models.
In fact, if we force the slope to be $n \propto 2m$ for
the four optimal data points (i.e., the squares in Figure~\ref{fig15})
and recompute the fit to find the intercept, we obtain $n = 2m - 2.067$,
which is close to the average of the two model-based curves derived
above from the central values.

\section{Conclusions}
\label{sec:conc}

The primary aim of this paper was to continue the process of narrowing
down the list of physical processes that have been proposed for
heating the solar corona.
This has been done by constructing LOS-integrated forward models of
coronal EUV/X-ray emission corresponding to a broad range
of suggested heating mechanisms, and comparing them with observations
from SDO/AIA and Hinode/XRT.
Our heating-rate prescription was flexible enough to account for
processes such as MHD waves, turbulence, footpoint braiding, and
magnetic reconnection (see Table~\ref{table01}).
This model of time-averaged coronal emission also included
up-to-date radiative losses, heat-flux gradients with ambipolar diffusion,
and DEM broadening from nanoflare-like heating intermittency.
The inclusion of ambipolar diffusion gives rise to loop
emission measures that decrease with increasing temperature from the
upper chromosphere to the transition region, which has been difficult
to achieve in time-steady models in the past.

From the analysis presented in Section~\ref{sec:results},
the region of parameter space for the heating-rate exponents that best
matched the data (i.e., $m \approx 1.5$ and $n \approx 2m-2$) was seen to
agree quite well with the predicted behavior of
{\em imbalanced MHD turbulence in closed coronal loops.}
Together with the successful use of imbalanced turbulence in global
MHD simulations of the solar wind \citep[e.g.,][]{Ri19,vdH22},
this agreement provides support for this process being a key
component of a complete theory of coronal heating.
Note that the combination of $m=1.5$ and $n=1$ produces a heating rate
that can be written as a parametric modification of the Kolmogorov
cascade rate,
\begin{equation}
  Q_{\rm heat} \, = \,
  \frac{\rho v_{\perp}^3}{\lambda_{\perp}}
  \left( \frac{\tau_{\rm ph}}{\tau_{\rm A}} \right)^{a}
\end{equation}
where, in this case, $a = 0.5$, and this parameter was specified as
$2-m$ in Equation~(\ref{eq:abdef}).
Note also that Table~\ref{table01} contains several examples of
models with $n=1$ and values of $a$ between about --1 and +2.
However, ours appears to be the first analytic prediction of this
scaling with $a = 0.5$.
Of course, this result does not come close to offering definitive proof
of imbalanced turbulence being the sole or primary driver of coronal
heating, but we believe it represents a meaningful step toward resolving
this long-standing problem.

The results presented above point to a slightly broader insight about
this kind of forward modeling: the use of a physically realistic scaling
for $Q_{\rm heat}$ (i.e., our $m$ and $n$ exponents) has advantages over
the purely empirical scalings discussed in, e.g.,
Section~\ref{sec:heating:other}.
Our chosen prescription ties together some of the overall plasma-parameter
scalings with the spatial dependence of the heating rate.
In other words, our choice of $m$ and $n$ exponents determines both
proportionalities such as $Q_{\rm heat} \propto B^a / L^b$ (as used in
many other papers) {\em and} the effective ``heating rate scale height''
$h$ (i.e., $Q_{\rm heat} \propto e^{-s/h}$, even though we do not
use this kind of exponential function).
However, if we had allowed $a$, $b$, and $h$ to vary completely
independently of one another, we would have explored some regions of
parameter space that do not correspond to self-consistent coronal heating
theories (see discussion in Section~\ref{sec:params:2nd}).

The remainder of this section discusses ways to improve these models
and obtain higher-confidence conclusions about the most likely
coronal heating mechanisms.
An important physical process that still needs to be included is
Taylor relaxation.
This process is the extraction of nonpotential magnetic energy from
twisted flux ropes, filaments, and sigmoid-shaped cores of active
regions \citep{T74,BP86,vB08,Ya18}.
Existing models of the heating rate for this process all depend on
the torsion parameter $\alpha$ or an equivalent dimensionless ``winding
number'' ($\alpha L$) that describes helical twists along a loop.
This quantity can be measured by comparing between observations and
nonpotential magnetic-field reconstructions \citep[e.g.,][]{Poi15,Xie17}.
Our neglect of Taylor relaxation may be the reason that the
most active cases that we studied had higher values of $T_{\rm norm}$
and lower values of both $m$ and $n$.

Modeling of this kind can also be improved in several other important ways:
\begin{enumerate}
\item
The PFSS extrapolation method is computationally efficient but also
notoriously inaccurate in reproducing the shapes of magnetic field
lines in the large-scale corona.
Well-studied improvements on this method include
the current-sheet source-surface method \citep{ZH95,PZ14},
the Schatten/McGregor current sheet method \citep{Mg08},
magnetofriction with solar wind expansion \citep{RY21},
and discrete semi-Lagrangian techniques \citep{Lw24}.
Several of these methods also take into account the acceleration of
solar wind along open field lines, which would also be useful to
build into our modeling framework as well.
\item
For active regions, it would be useful to reexamine our treatment of
the magnetic suppression of convective $v_{\perp}$ motions.
The quenching parameterization we used \citep[from][]{Sj04} can now be
validated by comparison with updated data from \citet{Ap25} and others.
\item
As summarized in Section~\ref{sec:plasma:cond}, our model of thermal
conduction can be made more realistic by including the effects of
flux-limited saturation for the active corona \citep[e.g.,][]{CD84}
and turbulent suppression \citep{Br19,AB25}.
\item
Our model of radiative cooling, discussed in Section~\ref{sec:plasma:rad},
could be improved by replacing the power-law fit from
Equation~(\ref{eq:radMa10}) by a function that more accurately
reflects the known shape of $\Lambda(T)$.
Also, since it is known that coronal abundances depend in some way
on magnetic field strength \citep[e.g.,][]{Ko16,To23},
it may be necessary to include some kind of dependence of $\Lambda$
on the local value of $B$.
\item
It is possible to make our solution for the time-steady loop pressure
$P(s)$ more self-consistent, both for the solution of the constants
$P_0$ and $T_{\rm max}$ (Section~\ref{sec:plasma:Ma10}) and for the
non-isothermal hydrostatic equilibrium (Section~\ref{sec:plasma:spatial}).
Note, however, that the present use of an isothermal temperature in
Equation~(\ref{eq:Psiso}) may be better justified than originally
assumed, since nonthermal (wave or turbulent) motions in the low corona
could compensate for the lower temperatures there and provide
something reasonably close to isothermal pressure support.
\item
Our interpretation of DEM broadening as originating from nanoflare
intermittency \citep[e.g.,][]{Ba16} was just one possible
physical explanation for this phenomenon.
This kind of broadening can arise from instabilities associated
with MHD turbulence \citep{VD18},
the existence of suprathermal electrons \citep{Dz15}, or
non-equilibrium ionization effects \citep{Shen23}.
These other processes are likely to produce broadening in the
temperature response functions with qualitatively different shapes
than those shown in Figure~\ref{fig04}.
Of course, true time dependence in the lower boundary conditions
\citep{Ms23} must also affect the time-averaged rate of coronal
heating in ways that the above models have not taken into account.
\end{enumerate}
Lastly, it will be important to include observational constraints
beyond the SDO/AIA and Hinode/XRT data used in this paper.
More complete context about coronal heating can be found by
incorporating cooler transition-region diagnostics
\citep[e.g.,][]{Bai23},
higher-energy emission \citep[e.g.,][]{Wr17}, and
data obtained at larger radii with the help of coronagraphic
occultation \citep[e.g.,][]{Ko06,GC20}.
Also, the spatial resolution of radio-frequency solar imaging has
improved substantially in recent years \citep[e.g.,][]{Fs21,Kn25}
which enables even more robust model/data comparisons.

\begin{acknowledgments}
The authors gratefully acknowledge
Daniel Mendoza, Garyfallia Strus, Aylecia Lattimer, and
Amy Winebarger for many valuable discussions,
and Janet Houser for indispensable assistance with SolarSoft.
The authors are also grateful to the anonymous referee for
many constructive suggestions that have improved this paper.
This work was supported by the
National Science Foundation (NSF) under grant 2300452, and by the
National Aeronautics and Space Administration (NASA)
under grant 80NSSC20K1319.
This work utilized solar magnetogram data produced collaboratively between
the Air Force Research Laboratory (AFRL), who developed the ADAPT model,
and the National Solar Observatory's Integrated Synoptic Program (NISP),
which operates the GONG telescope network.
Data from SDO is used courtesy of NASA/SDO and the AIA, EVE, and HMI
science teams.
Data from Hinode/XRT is used courtesy of ISAS/JAXA, NASA, and
the Smithsonian Astrophysical Observatory.
Hinode is a Japanese mission developed and launched by ISAS/JAXA, with
NAOJ as domestic partner and NASA and STFC (UK) as international partners,
and it is operated by these agencies in co-operation with ESA and
NSC (Norway).
This research made extensive use of NASA's Astrophysics Data System (ADS).
\end{acknowledgments}

\end{document}